%% file: main.tex
\documentclass[opre,nonblindrev]{informs3} 

\OneAndAHalfSpacedXI 

\usepackage{endnotes}
\let\footnote=\endnote

\newcommand{\hide}[1]{}
\usepackage{natbib}
\NatBibNumeric
\def\bibfont{\small}%
\bibpunct[, ]{[}{]}{,}{n}{}{,}%
\usepackage{flushend}

\usepackage{url}
\makeatletter
\g@addto@macro{\UrlBreaks}{\UrlOrds}
\makeatother
\usepackage{latexsym}
\usepackage{amsmath}
\usepackage[normalem]{ulem}
\usepackage{amssymb}
\usepackage{color}
\usepackage[sort]{cite}
\usepackage{epstopdf}
\usepackage{amsfonts}
\usepackage{mathrsfs}
\usepackage{aliascnt}
\usepackage{enumerate}
\usepackage{graphicx}
\usepackage{hyperref}
\usepackage{algpseudocode}
\usepackage{algorithm}
\usepackage{mathtools}
\usepackage{ifthen}
\newboolean{showcomments}
\setboolean{showcomments}{true}
\usepackage[normalem]{ulem}

\theoremstyle{plain}

\input{macros}

\input{notation}

\newcommand{\rone}[1]{\textcolor{black}{#1}}
\newcommand{\sam}[1]{\textcolor{black}{#1}}
\newcommand{\adam}[1]{\textcolor{black}{}}
\newcommand{\eb}[1]{\textcolor{black}{#1}}
\newcommand{\ebc}[1]{\textcolor{brown}{}}

\usepackage[noabbrev,capitalize]{cleveref}

\def\EMAIL#1{\href{mailto:#1}{#1}}

\usepackage{natbib}
 \bibpunct[, ]{(}{)}{,}{a}{}{,}%
 \def\bibfont{\small}%
\TheoremsNumberedThrough     
\ECRepeatTheorems

\EquationsNumberedThrough    

\MANUSCRIPTNO{OPRE-2019-03-125} 

\begin{document}


\RUNAUTHOR{Pang et al.} 

\RUNTITLE{Transparency and Control in Platforms for Networked Markets}

\TITLE{Transparency and Control in Platforms for Networked Markets}

\ARTICLEAUTHORS{%
	\AUTHOR{John Pang}
	\AFF{California Institute of Technology, \EMAIL{johnpzf@gmail.com}}
	\AUTHOR{Weixuan Lin}
	\AFF{Cornell University, \EMAIL{wl476@cornell.edu}}
	\AUTHOR{Hu Fu}
	\AFF{University of British Columbia, \EMAIL{hufu@cs.ubc.ca}}
	\AUTHOR{Jack Kleeman}
	\AFF{University of Cambridge, \EMAIL{jackkleeman@gmail.com}}
	\AUTHOR{Eilyan Bitar}
	\AFF{Cornell University, \EMAIL{eyb5@cornell.edu}}
	\AUTHOR{Adam Wierman}
	\AFF{California Institute of Technology, \EMAIL{adamw@caltech.edu}}
} 
\input{abstract}


\MSCCLASS{}
\ORMSCLASS{Primary: Games/Group Decisions; secondary: Networks/Graphs }
\HISTORY{}

\maketitle
\section{Introduction} \label{intro}
\input{intro}

\section{Preliminaries}
\input{prelims}

\section{Open Access Platforms}\label{sec:oap}
\input{openaccess}

\section{Controlled Allocation Platforms} \label{sec:cap}
\input{controlledplatform}

\section{Discriminatory Access Platforms} \label{sec:dap}
\input{discplatform}

\section{Conclusion and Future Work}
\input{conclusion}

\begin{APPENDICES}
	\input{appendix}

\end{APPENDICES}

\section*{Acknowledgments.}
We would like to thank Yu Su, and the two anonymous reviewers, for their helpful and constructive feedback in improving this work. 

\bibliographystyle{informs2014} 
\bibliography{biblio.bib,desmond.bib} 

\end{document}

%% file: macros.tex
\usepackage[usenames,dvipsnames,svgnames,table]{xcolor}

%% file: notation.tex
\newcommand{\Xcomment}[1]{{}}

\newcommand{\dd}{\mathrm{d}}  

\newcommand{\eps}{\epsilon}

\newcommand{\noaccents}[1]{#1}

\newcommand{\newagentvar}[3][\noaccents]{%
\expandafter\newcommand\expandafter{\csname #2\endcsname}{#1{#3}}%
\expandafter\newcommand\expandafter{\csname #2s\endcsname}{#1{\boldsymbol{#3}}}%
\expandafter\newcommand\expandafter{\csname #2smi\endcsname}[1][i]{#1{\boldsymbol{#3}}_{-##1}}%
\expandafter\newcommand\expandafter{\csname #2i\endcsname}[1][i]{#1{#3}_{##1}}%
\expandafter\newcommand\expandafter{\csname #2ith\endcsname}[1][i]{#1{#3}_{(##1)}}%
}

\newcommand{\profit}{\pi}
\newcommand{\price}{p}
\newcommand{\demlim}{\alpha}
\newcommand{\demrate}{\beta}
\newcommand{\supply}{s}
\newcommand{\demand}{d}
\newcommand{\production}{q}
\newcommand{\cost}{C}

\DeclareMathOperator{\SW}{SW}
\DeclareMathOperator{\REV}{REV}
\DeclareMathOperator{\CS}{CS}

\newcommand{\Rset}{\mathbb{R}}

\newcommand{\Ecal}{{\cal E}}

\newcommand{\Lcal}{{\cal L}}

\newcommand{\Ocal}{{\cal O}}

\newcommand{\Qcal}{{\cal Q}}

\makeatletter
\newcommand{\hufu}[1]{\textcolor{red}{[Hu\@ifnotempty{#1}{: #1}]}}
\makeatother

%% file: abstract.tex
\ABSTRACT{%
In this paper, we analyze the worst case efficiency loss of online platform designs under a networked Cournot competition model. Inspired by some of the largest platforms in operation today, the platform designs that we consider examine the trade-off between  transparency and control. Our results show that open access designs incentivize increased production towards perfectly competitive levels and limit efficiency loss, while controlled allocation designs lead to producer-platform incentive misalignment, resulting in low participation and unbounded efficiency loss. We also show that discriminatory access designs balance transparency and control, achieving the best of both worlds by maintaining high participation rates while limiting efficiency loss. 
}%

%% file: intro.tex
The recent emergence of online platforms has driven tremendous innovation in the design and operation of two-sided marketplaces. Platforms, unlike traditional retailers,  do not typically produce or store physical goods, but instead match producers to consumers and set prices algorithmically. For example, the majority of ride-sharing platforms (today) do not employ their drivers, but instead match available drivers to riders using mobile apps. 
Apart from ride-sharing, there are many other platforms that connect producers to consumers in innovative ways to improve both the provisioning of supply and the fulfillment of demand, which has led  to enormous efficiency gains and explosive platform growth across diverse marketplaces. \citet{evans2016rise} estimate the market value of platform companies to be over 4.3 trillion USD, representing a significant fraction of the  56.8 trillion USD aggregate  market value the largest $2000$ companies globally  \citep{forbesreport}.

\eb{There is a growing literature on the subject of platform design that seeks to understand how matching, access control, and pricing impact the efficiency of platform markets.}
In a seminal paper on two-sided markets, \citet{rochet2003platform} compare market outcomes stemming from different governance structures, e.g., profit-maximizing versus not-for-profit organizations. Using a  similar two-sided market model, \citet{armstrong2006competition} compares platforms with  three different access control designs: monopolistic platforms, competing platforms where each user 
can only choose one platform, and bottleneck platforms where \rone{only one side of the market (e.g., the consumer side)} can engage with multiple platforms.
More recently, \citet{akbarpour2017thickness} have shown under a model of dynamic matching that planners can improve overall utility with information on when agents depart as waiting patiently for a thicker market on both sides often allow for better matching, whenever it is feasible. 
The impact of algorithmic pricing on  real-world markets has also been  studied in recent years  \citep{chawla2007algorithmic, chawla2010multi,li2012automated}. Lately, \citet{dinerstein2018consumer} have focused on balancing trade-offs around search friction between conflicting designs, e.g., how much to guide consumers and reduce search friction while incentivizing price competition between sellers. 

\emph{In this paper, we investigate the trade-off between  transparency and control that arises in the design of access and allocation control mechanisms in platforms.}
Present-day platforms exhibit a diverse array of access and allocation control designs.
For example, eBay adopts an \emph{open access} design that displays the offers from all producers to the customers. This design is intended for lowering producer entry cost and boosting competition, but may end up suffering from high consumer search cost. Uber, on the other hand, adopts a \emph{controlled allocation} design, in the sense that it only allows drivers to accept or decline the ride that is assigned from the platform. Lastly, Amazon employs a \emph{discriminatory access} design by directing its customers to its pre-selected ``Buybox" producers. While such an approach  serves to limit consumer search costs,  it may  lack transparency due to the ambiguity in the criterion for ``Buybox eligibility" (cf. \citet{chen2016empirical}).

There are a variety of important open questions related to the design of access and allocation control mechanisms for platforms. First, what is the worst case efficiency loss of platforms under these designs? Second, what is the impact of allocation control on worst case efficiency loss
, and are there any unintended incentives for strategic behavior that might decrease market efficiency? Lastly, is there a sweet spot between open access and controlled allocation mechanisms that balances transparency and control, while limiting efficiency loss? 

\subsection{Our Contributions}

In this paper, we model and analyze three different platform designs: (i) open access, (ii) controlled allocation and (iii) discriminatory access designs. \eb{Using a networked Cournot competition model as the basis for our analysis,  we provide worst case efficiency loss bounds for each design.}
\eb{At the heart of the proposed model is a network (represented as a bipartite graph) that  specifies connections between firms and markets, where each firm competes \'{a} la Cournot in each of the markets that is has access to. In open access platform designs, all firms have access to all markets, resulting a complete bipartite graph. In discriminatory access designs, the platform determines the structure of network, i.e., the collection of markets that each firm is given access to.} \sam{In contrast, in controlled allocation platform designs, firms choose their individual supply levels, and the platform determines the allocation of the aggregate supply generated by the firms to the different
markets.} \eb{In this paper, we aim to shed light on the various trade-offs that emerge between efficiency loss, transparency, and control across these different platform designs.}

The first platform design we consider is the open access design (Section \ref{sec:oap}), exemplified by online marketplaces like eBay and Etsy. The advantages of an open access platform design are its fairness, transparency, low producer entry cost, and the increased  market competitiveness driven by such attributes. 
\sam{We prove that  open access  platform designs ensure  that at least half of the socially efficient demand level in each market is fulfilled at Nash equilibrium.}
%
 We also show that the social welfare levels achieved  under open access platform designs at equilibrium are no less than  $2/3$ of the efficient social welfare in the worst case. (Theorem \ref{thm:PoA_asym}). 
\eb{In proving this result, we employ an intermediary argument---originally proposed by  \citet{johari2005efficiency} in the context single-market Cournot models---which shows that, among all convex producer cost functions,  \emph{linear} cost functions result in the greatest loss of efficiency at equilibrium. Building on this argument, we are also able to establish bounds that reflect the impact of asymmetry in  producer cost functions on  worst-case efficiency loss at equilibrium.}
\sam{Together, these results show that open access platform designs promote sales volume across the different markets,} \eb{while limiting efficiency loss at equilibrium.}

\eb{While open access platforms are shown to exhibit bounded efficiency loss, the welfare loss stemming from the ``misallocation'' of supply across the different markets can be substantial.
As an alternative to open access platforms, we investigate a  controlled allocation platform design (Section \ref{sec:cap}), where the platform determines the allocation of supply across the different markets given quantity offers from the different producers.} 
%
%
For example, in ride-sharing platforms, drivers typically  decide when to drive, \eb{while the platform decides who they pick up.} 
\sam{As one of our main results, we show that the worst-case efficiency loss incurred by controlled allocation platforms grows at least linearly in the number of markets for a large family of supply allocation mechanisms (Theorem \ref{thm:generalcap}).} Perhaps surprisingly, this lower bound on the worst-case efficiency loss is shown to hold even in settings where the platform allocates supply across the different markets to maximize social welfare (Theorem \ref{thm:ineff}). The efficiency loss incurred by controlled allocation platform designs suggests that explicit allocation control may incentivize a reduction in production levels to such an extent that the efficient allocation of supply across the different markets cannot offset the efficiency loss associated with reduced production. 

\sam{As an alternative to controlled allocation platforms, we study a discriminatory access platform design where the platform indirectly controls the allocation of supply to consumers by restricting the set of markets that each seller has access to.}
%
%
%
%
\eb{In this context, the set of markets that each firm has access to is determined by the platform in order to maximize  social welfare  at  the Nash equilibrium induced by the resulting  networked Cournot game.}
%
%
\eb{In Theorem \ref{thm:discriminatory_PoA}, we show that discriminatory access platforms improve upon the worst-case efficiency loss of open access and controlled allocation platforms, resulting in Nash equilibria whose  social welfare  is no less than 3/4 of the efficient social welfare in the worst case.}
\sam{Given a simplifying assumption on the linearity of producers' cost functions, we  provide a greedy algorithm that is guaranteed to generate the optimal network (access control) structure, which  maximizes social welfare at Nash equilibrium (Theorem \ref{thm:greedy-alg}).}
\eb{Our analysis also sheds light on the role of producer cost asymmetry in determining  the optimal network structure. If producers have identical cost functions, then the optimal network structure is open access.  
If, however, there is significant asymmetry between producers' cost functions, then the optimal network structure may bar certain firms from participating in certain markets in order to preserve the ``competitiveness'' of those markets.}

\subsection{Related Literature}

\eb{In this paper, we examine the role  of access and allocation control on the worst-case efficiency loss in online platforms where producers compete \`{a} la Cournot.}
\sam{Our work contributes to two major streams of literature relate to (i) the design and operation of two-sided platforms, and (ii) the  analysis of competition in networked markets.}




\subsection*{Platform Design and Operation}


The recent growth of online platforms has led researchers to focus on identifying design features common to successful platforms. Earlier works in this area started by introducing different models of two-sided platform markets. \rone{The renowned work of \citet{rochet2003platform} considered a model of platform competition in two-sided markets with network externalities, e.g., markets where the buyer surplus depends on the number of sellers.}
\sam{In that work, the authors leverage the platform competition model to quantify how the consumer and producer surplus is affected by the market clearing mechanism in the platform.}
\sam{In \citet{evans2005industrial}, the authors provide additional applications of Rochet and Torile's model on the antitrust regulation of two-sided markets.}
Using a different model, \citet{armstrong2006competition} considers the presence of important cross-group or indirect network effects between the customer groups participating on the platform. Other definitions and models are found in many other work, such as \citet{hagiu2015multi}, which compared the choice for a firm between operating within a multi-sided platform setting and operating independently via vertical integration. 
A good summary of these earlier works can be found in \citet{rysman2009economics}. 
Our focus is on access and allocation control designs for these platforms.  

\subsubsection*{Open Access Platforms.}
Open access is touted in \citet{boudreau2010open} as a way to increase competition among participants \rone{in mobile device platforms}. In \citet{parker2017innovation}, the open access design of \sam{online platforms for developers} is shown to promote fairness and \sam{innovation}.
The classical example studied for open access \rone{for retail markets} is the online marketplace eBay.
\sam{The study of eBay in \citet{gross2003balances,hui2016reputation} reveals that the success of open access platforms will, in general, depend heavily on the reputation of sellers and the platform regulation policies (e.g., seller ratings and buyer protections).}



\subsubsection*{Controlled Allocation Platforms.} 
In contrast to the open access approach, some platforms (e.g., Uber) may choose to directly determine the allocation of supply to the consumers. 
\sam{These platforms, however, might be susceptible to the manipulation from the suppliers.}
For example, it is also well known and reported in \citet{chen2017thrown} that drivers on ride-sharing platforms collaborate and reduce their supply to cause price surge and demand spikes in the system. 
%
\sam{Several papers have investigated pricing and allocation policies that might limit the price manipulation from suppliers. Examples of such work include the study on a spatio-temporal pricing mechanism in \citet{ma2018spatio} and the study on admission control policies in \citet{afeche2018ride}.
Apart from the potential market manipulation from suppliers, 
other major challenges in managing controlled allocation platforms include the difficulty in solving the supply allocation problem, and the potential insufficiency of supply that results from the competition with rival platforms. We refer the readers to \citet{alijani2017two, DBLP:journals/corr/Banerjee0L16, scheiber2017uber,lu2018surge,fang2018loyalty,banerjee2015pricing} for recent advances.
}



\subsubsection*{Discriminatory Access Platforms.}

Another possible way to improve upon open access platforms is through discriminatory access. This is studied in \citet{banerjee2017segmenting,kanoria2017facilitating,akbarpour2017thickness}, where the platform restricts the set of markets that producers are allowed to bid in to improve social welfare. 
A typical example of discriminatory access platforms is Amazon's Buybox studied in \citet{chen2016empirical}, where Amazon highlights one seller for each item. Another example of discriminatory access is Airbnb's Superhost program, which highlights certain hosts through badges. \sam{In \citet{liang2017superhost}, it was shown that the Superhost program vastly increases Airbnb's revenue.}
\sam{The majority of prior work on discriminatory access platforms, e.g., \citet{chawla2007algorithmic,chawla2010multi,chen2017observing}, has focused on the effect of algorithmic pricing on the efficiency of the platform. In this paper, we instead examine the reduction in worst-csae efficiency loss (relative to open access designs) that can be attained through discriminatory access.}


\subsubsection*{Alternative Design Factors.} Besides access and allocation control, work in this area has also covered a variety of possible design factors, including pricing \citep{weyl2010price}, competition  \citep{armstrong2006competition}, reputation \citep{nosko2015limits,tadelis2016reputation,luca2017designing}, and thickness or volume \citep{ashlagi2018matching}. 
Recent empirical studies in \citet{einav2015assessing} and \citet{bimpikis2016spatial} also reveal significant price dispersion in online e-commerce and ridesharing marketplaces, 
causing platforms to differentiate products in order to create distinct consumer markets as in \citet{dinerstein2018consumer}. In particular, these results highlight the need to study platforms in using models of \emph{networked competition}.

\subsection*{Networked Competition}
Models of networked competition aim to capture the effects of network constraints on strategic interactions between firms. They are prevalent in industries which have distinct geographical markets with significant price differentiation. 
\sam{Depending on the model's assumption on the decision variables of the producers, these models can be classified into the networked Bertrand model \citep{chawla2008bertrand, guzman2011price, anshelevich2015price}, the networked Cournot model \citep{ilkilic2009cournot, bimpikis2014cournot, abolhassani2014network,  xu2017efficiency,le2018efficiency}, the networked supply function model 
\citep{berry1999understanding, hobbs2000strategic, liu2007impacts, wilson2008supply, holmberg2014supply, lin2019structural}, and other networked bargaining games where agents can trade via bilateral contracts over a network that determines the set of feasible trades (cf.  \citet{AbreuManea2012, Elliott2015, Nava2015, nguyen2015coalitional}).}
These networked competition models have applications in many areas, e.g., electricity markets in \citet{bose2014role} and demand-side management in smart grids in \citet{motalleb2017game}.


Our work relies on a  model of networked Cournot competition. 
\sam{Compared with other models of networked competition, the advantage of the networked Cournot model is its  balance between explanatory power and tractability of analysis. Specifically, Cournot models are able to capture the potential exercise of market power from producers via the withholding of production. The Bertrand model, in contrast, assumes that each producer is willing to produce any amount of good at its offered price, and might overestimate the intensity of competition between producers (cf. \citet{yao2008modeling}). On the other hand, although one might obtain additional flexibility in describing the price-quantity pair that each firm offers via the use of supply function models, the equilibrium analysis of such models are, in general, intractable if a firm has access to multiple markets with heterogeneous demand functions (cf. \citet{wilson2008supply}).  }

The vast majority of the literature on networked Cournot competition, e.g., \citet{bimpikis2014cournot, abolhassani2014network, ilkilic2009cournot}, focuses on characterizing and computing Nash equilibria in the absence of capacity constraints. Recent work from \citet{alsabahcapacity} focuses on a scenario with capacity constraints. This paper is most closely related to \citet{johari2005efficiency}, which provide bounds on the worst case efficiency loss of Cournot games. \citet{cai2017role} shows how the mechanism of market clearing (which is determined via the choice of the system operator's objective function) might affect the efficiency of the platform under network constraints. Our work focuses on platform design through access and allocation control under the networked competition model. We extend some key results from \citet{johari2005efficiency} on worst case efficiency loss while presenting new results on how a platform might reduce the worst-case efficiency loss at Nash equilibrium via appropriate access control.

%% file: prelims.tex
\label{sec:prelims}

\begin{figure}[t]
    \centering
    \includegraphics[width=0.85\textwidth]{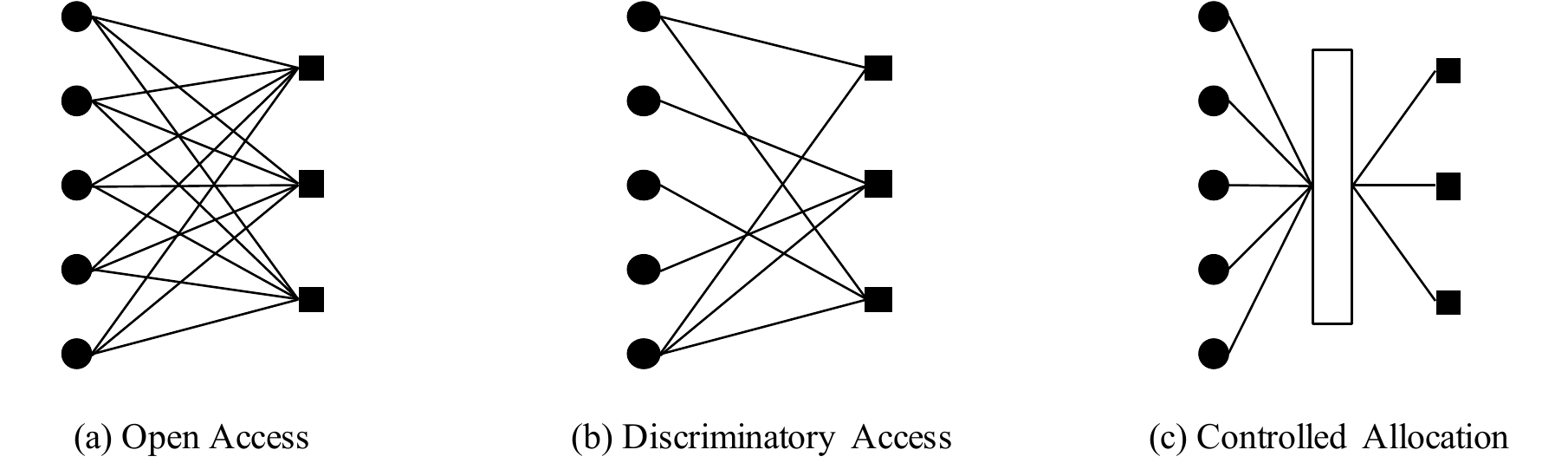}
    \caption{\eb{From left: (a) open access platforms allow all firms access to all markets, (b) discriminatory access platforms restrict the set of markets that each firm has access to, and (c) controlled allocation platforms determine the allocation of supply to markets given quantity offers from the firms. All edges are directed from firms (circular nodes) to markets (square nodes), and removed for clarity.} }
    \label{fig:platforms}
\end{figure}

\eb{In this paper, we adopt a model of platform competition in which firms compete  \'{a} la Cournot to produce a homogeneous product in multiple markets, where the platform serves to mediate the interactions between firms and markets.  Within this framework, we examine three distinct approaches to platform control: \emph{open access}, \emph{discriminatory access}, and \emph{controlled allocation} designs, as illustrated in Fig. \ref{fig:platforms}. In open access platforms, all firms are given access to all markets. In discriminatory access designs, the platform  explicitly controls (restricts) the set of markets that each firm is given access to. In controlled allocation designs, the platform determines the allocation of supply across the different markets given quantity offers from the different firms.} 

\eb{At the heart of our models for open and discriminatory access platforms is a directed bipartite graph $(F, M, \mathcal{E})$ that specifies the set of markets that each firm is given access to.}
Here, $F := \{1, \dots, n\}$ denotes the set  of $n$ firms,   $M := \{1, \dots, m\}$ denotes the set  of $m$ markets, and 
$\mathcal{E} \subseteq F \times M$ denotes the set of  directed edges connecting firms to markets. That is to say, $(i,j)\in \mathcal{E}$ if and only if  firm $i \in F$ has access to market $j \in M$. In what follows, we develop mathematical models for each of these three classes of platform designs.

\subsection{Open Access Platforms}
\eb{In open access platforms, all firms are given access to all markets. Examples of such platforms 
include eBay and Etsy, where every customer is shown every retailer that sells the item she desires.
Mathematically, open access platforms can be described according to networks  $(F, M, \mathcal{E})$ with a complete set of directed edges from firms to markets given by $\Ecal = F \times M$.}

\eb{We now  introduce the various components that form the basis for our model of networked Cournot competition in open access platforms.} We discuss how to adapt this model to accommodate controlled allocation and discriminatory access platform designs in the following sections.



\emph{Producer model:} \label{sec:producer}
In open access platforms, each firm determines how much it will produce in each market. Accordingly, we let $\production_{ij} \in \Rset_+$  denote the quantity produced by firm $i$ in market $j$, and let $\production_i :=  ( \production_{i1}, \dots, \production_{im} ) \in \mathbb{R}_+^m$ denote the supply profile associated with firm $i$. We define the set of feasible supply profiles from firm $i$ as
$$\Qcal_i (F \times M) := \left\{ x \in \mathbb{R}_+^m \, \left| \, x_j = 0, \ \forall \ (i, j ) \notin F \times M \right.  \right\}.$$ 
We  denote the supply profile across all producers by $q:=(q_1, \dots, q_n)$, and define the corresponding set of feasible supply profiles from all firms  by $\mathcal{Q} (F \times M) := \prod_{i = 1}^n \mathcal{Q}_i (F \times M)$. We let $\supply_i \in \mathbb{R}_+$ denote the aggregate quantity produced by each firm $i \in F$. It is given by
\begin{equation}
\supply_i := \sum_{j=1}^m \production_{ij}. 
\end{equation}
\eb{The cost incurred by each firm $i$ to produce an aggregate quantity $s_i$ is defined according to $\cost_i (\supply_i)$, where   $\cost_i: \mathbb{R} \rightarrow \mathbb{R}_+$ denotes each firm's  production cost function.}  We let $C := (C_1, \dots, C_n)$ denote the cost function profile across all firms, \eb{which is treated as common knowledge.} \eb{The cost functions considered in this paper are assumed to satisfy the following assumption.
\begin{assumption}  \label{ass:cost}
Each firm's  cost function $\cost_i$ is assumed to be convex, differentiable on $(0,\infty)$, and satisfy $\cost_i(x)=0$ for all $x\le 0$. 
\end{assumption}}

\emph{Demand model:}
As is standard in Cournot models of competition, we model price formation in each market according to a commonly known inverse demand function.
Similar to \citet{bimpikis2014cournot}, we restrict our attention to affine inverse demand functions throughout this paper. Specifically, the price in each market $j \in M$ is determined according to 
$$\price_j(\demand_j):=\demlim_j - \demrate_j \demand_j,$$
where $\demand_j :=  \sum_{i=1}^n \production_{ij}$  denotes the aggregate quantity supplied to market $j$. The parameters $(\alpha_j, \beta_j)$ are  assumed to be strictly positive  for all $j \in M$.





\emph{Social welfare:}
We measure the performance of a platform according to the \textit{social welfare} that it generates at equilibrium. The maximization of social welfare benefits both buyers and sellers,  promoting  platform growth in the long run. 
Social welfare is defined as aggregate consumer utility 
minus the aggregate production cost. Specifically, the social welfare induced by a particular supply profile $\production$  is given by
	\begin{align}
	\label{eq:welfare-def}
	\SW(\production,\cost) := \sum_{j=1}^m \int_0^{\demand_j}\price_j(z) \: \dd z - \sum_{i=1}^n \cost_i(\supply_i),
	\end{align}
where $\supply_i :=  \sum_{j =1}^m q_{ij}$ denotes the aggregate supply of firm $i \in F$  and $\demand_j := \sum_{i=1}^n q_{ij}$ denotes the aggregate demand in market $j \in M$. \eb{We have made explicit the dependence of social welfare on the cost profile $\cost$  to facilitate later discussions and proofs, where we quantify the dependence of social welfare on the underlying cost profile.} The \emph{efficient social welfare} associated with a cost  profile $\cost$ and the edge set $F\times M$ is defined as:
\begin{equation}
\SW^*(F\times M,\cost) := \sup_{\production \in \mathcal{Q}(F\times M)}~ \SW(q,\cost).
\end{equation}
Under the the paper's standing assumptions, it is straightforward to show that the above supremum is achieved, and that the set of efficient supply profiles is non-empty. In Section \ref{sec:disc_acc}, we adapt the definition of efficient social welfare to the setting of discriminatory access platforms.

\subsection{Controlled Allocation Platforms}
\eb{In the family of controlled allocation platform designs that we consider, firms choose their individual production levels, while the platform determines the allocation of the resulting aggregate supply across the different markets to maximize a combination of aggregate consumer surplus and  producer revenues.}
In ridesharing platforms, for example, drivers typically determine the number of hours that they work, \eb{while the platform determines the specific rides that they're matched to, and, as a result, the geographic regions (markets) that they are routed to.} \rone{In reality, drivers can also influence (in part) the market that they serve by choosing to go offline and reroute to different regions.}\ebc{[Eilyan says: This is not a perfect example of controlled allocation platforms, as I would not necessarily equate the matching of drivers to riders with  ``market allocation''. Rather, drivers can influence (in part) the market that they serve by choosing to go offline and reroute to different regions.]} 
 

More formally, in a controlled allocation platform, 
each firm $i \in F$ chooses its aggregate production quantity $s_i \in \mathbb{R}_+$, incurring a production cost $C_i(s_i)$. \eb{The platform determines the aggregate quantity $d_j \in \mathbb{R}_+$ supplied to each market $j \in M$, subject to the constraint $\sum_{j=1}^m d_j \leq Q$, \sam{where $Q := \sum_{i=1}^n s_i$.}
We consider a family of market clearing mechanisms where the platform determines the allocation of supply to maximize a convex combination of consumer surplus and producer revenue, given by
\begin{align*}
    \textrm{OBJ}(d; \lambda) := \lambda \cdot \CS(d) + (1- \lambda) \cdot \REV(d),
\end{align*}
where $\textrm{CS}: \Rset_+^m \rightarrow \Rset$ and $\textrm{REV}: \Rset_+^m \rightarrow \Rset$ 
denote the  aggregate consumer surplus and aggregate producer revenue functions, respectively. They are given by}
\rone{
\begin{align*}
    \CS(d) &:= \sum_{j=1}^m \int_0^{\demand_j}\price_j(z)\textrm{dz} - \demand_j\price_j(\demand_j),\\
    \REV(d) &:= \sum_{j=1}^m  \demand_j\price_j(\demand_j).
\end{align*}} 
\eb{The parameter $\lambda \in [0,1]$ can be specified by the platform to tradeoff the competing interests of  consumers and  producers. For example, setting $\lambda = 1/2$ results in a platform objective function $\textrm{OBJ}(d; 1/2)$ that corresponds to aggregate consumer utility.} 
\rone{Given a particular choice of $\lambda \in [0,1]$, the controlled allocation platform performs the following optimization to determine the allocation of supply to each market:}
\eb{
\begin{align}
\label{opt:cont} \textrm{maximize} ~&~ \textrm{OBJ}(d;\lambda)\\
\nonumber \textrm{subject to}   ~&~ \sum_{j=1}^m d_j =  Q,
\end{align}
where $Q \in \mathbb{R}_+$ is determined by the firms, and $d \in \mathbb{R}_+^m$ is the platform's decision variable.} \eb{Given an efficient allocation $d$, we  assumed that firms are remunerated according to a \emph{uniform pricing mechanism} that is also \emph{budget balanced}. The unique pricing mechanism $p: \mathbb{R}_+^m \rightarrow \mathbb{R}$ that satisfies these criteria is given by  
\begin{equation}\label{eq:capprice}
p\left(  d\right) := \frac{\sum_{j=1}^m d_j p_j(d_j)}{\sum_{j=1}^m d_j}.
\end{equation}
Controlled allocation platforms, being defined in this manner, can be interpreted as giving rise to a multi-leader, single-follower Stackelberg game, where  firms choose their production quantities simultaneously, in anticipation that price will be determined according to \eqref{opt:cont}-\eqref{eq:capprice}. In specifying the firms' payoff functions in Section \ref{sec:cap}, it will be convenient to express the optimal solution to the platform's optimization problem \eqref{opt:cont} in the form of an \emph{allocation function} $A_\lambda:\mathbb{R}_+ \rightarrow \mathbb{R}^m_+$, which maps the aggregate production quantity $Q = \sum_{i\in F} s_i$  to an allocation $d$. We have parameterized the allocation function in terms of $\lambda$ to emphasize its dependence on the specific  objection function being optimized by the platform.}


\eb{Given a supply profile $s \in \Rset_+^n$, one can express the \emph{social welfare} derived under the controlled allocation platform according to
\begin{align}
\label{eq:welfare-def-stack}
\SW(\supply, \lambda) := \sum_{j=1}^m \int_0^{d_j} \price_j(z) \: \dd z - \sum_{i=1}^n \cost_i(\supply_i),
\end{align}
where $d = A_\lambda \big(\sum_{i \in F} \supply_i \big)$.
}
\rone{Here, we have made explicit the dependence of social welfare on the platform objective parameter $\lambda$ to facilitate the presentation of our results in the sequel.} \eb{We also note that, while we have reused the notation $\SW(\cdot, \cdot)$  in  defining  the  social welfare function associated with each platform design, the specific platform being considered when examining social welfare will always be clear from the surrounding context.}

\subsection{Discriminatory access platforms} \label{sec:disc_acc}
A discriminatory access platform restricts the set of markets that can be accessed by each firm in the form of an edge set $\Ecal\subseteq F \times M$. This restriction may be designed to improve producer surplus, consumer surplus, or social welfare. An example of a discriminatory access platform design is Amazon's Buy Box, where Amazon chooses a default seller based on a score that is based on product pricing, availability, fulfillment capabilities, and customer service records. Operationally, these restrictions prevent certain firms from accessing certain markets. We encode these restrictions in form of sparsity constraints on the supply profile of each firm. Formally, each firm~$i$ determines the quantity~$\production_{ij}$ that it supplies to each market~$j$, subject to the restriction 
that $\production_{ij} = 0$ for all $(i, j) \notin \Ecal$. We denote the set of feasible supply profiles associated with each firm $i \in F$ by
$$\Qcal_i (\Ecal) := \left\{ x \in \mathbb{R}_+^m \, \left| \, x_j = 0, \ \forall \ (i, j ) \notin \Ecal \right.  \right\}.$$
We denote the set of feasible supply profiles across all firms by $\mathcal{Q} (\Ecal) := \prod_{i = 1}^n \mathcal{Q}_i (\Ecal)$. In studying  discriminatory access platform designs, we assume that the supplier cost functions and  demand functions are identical to those considered in the context of open access platform designs.

We define the \emph{efficient social welfare} 
associated with an edge set $\Ecal$ and a cost function profile $\cost$ as
\begin{equation}
\SW^*(\Ecal, \cost) := \sup_{\production \in \Qcal(\Ecal)}~ \SW(q,\cost).
\end{equation}
A supply profile $\production \in \Qcal(\Ecal)$ is said to be efficient with respect to a given edge set $\mathcal{E}$ if it satisfies $\SW(\production,\cost) = \SW^*(\Ecal,\cost)$. It is straightforward to verify that the above supremum is  attained, and that the set of efficient supply profiles is non-empty. 


\subsection{Networked Competition Model}\label{subsec:ncm}
We now introduce the game theoretic models that we use to characterize market equilibrium in each of the  three platform designs considered in this paper.

\subsubsection{Competition under open access and discriminatory access platforms:} 
We consider profit maximizing firms, where the payoff function $\pi_i$ for each firm $i \in F$ is defined as
\begin{equation}
\profit_i(\production_i,\production_{-i})  := \sum_{j=1}^m \production_{ij}\price_j(\demand_j)
- \cost_i (\supply_i).
\end{equation}
Here, $q_i$ denotes the supply profile of firm $i$, and $\production_{-i}=(q_1, .., q_{i-1}, q_{i+1}, .., q_n) $ denotes the supply profiles of all firms not including firm $i$, and $d_j := \sum_{i=1}^n q_{ij}$ denotes the aggregate quantity consumed in each market  $j \in M$.


\eb{Given an edge set $\Ecal \subseteq F \times M$,} the triple $(F,\Qcal(\Ecal),\profit)$ defines a normal-form game, which we refer to as a \emph{networked Cournot game}. \eb{Depending the particular choice of edge set, the \emph{networked Cournot game} can be used to describe competition in either open access platform designs ($\Ecal = F \times M$), or discriminatory access platform designs ($\Ecal \subset F \times M$).}
%
We describe the \eb{family of  stable outcomes} associated with each  networked Cournot game using the concept of Nash equilibrium.

\begin{definition}[Nash Equilibrium]
	\emph{A supply profile $\production \in \mathcal{Q}(\Ecal)$ constitutes a \emph{pure strategy Nash equilibrium} of the game $(F,\mathcal{Q}(\Ecal),\profit)$ if for every firm $i\in F$, it holds that $\profit_i(\production_i,\production_{-i}) \ge \profit_i(\overline{\production}_i,\production_{-i})$ for all  $\overline{\production}_i \in \mathcal{Q}_i (\Ecal)$.} 
\end{definition}

Under the assumptions of convex cost functions and affine inverse demand functions, \citet{abolhassani2014network} have previously shown that the networked Cournot game is an ordinal potential game. Additionally, it admits a unique Nash equilibrium, which can be characterized as the unique optimal solution to a particular convex program.
We summarize the results of \citet{abolhassani2014network} in the following lemma.

\begin{lemma}\label{lem:NE} \citep{abolhassani2014network}
	The game $(F, \Qcal(\Ecal), \pi)$ admits a unique Nash equilibrium $q^{\rm{NE}}(\Ecal) \in \Qcal(\Ecal)$ that is the unique optimal solution to the following convex program:
	\begin{equation}
	\begin{alignedat}{8}
	&\underset{q \in \Qcal(\Ecal)}{\text{\rm maximize}} \quad && \SW (q, C) - \sum_{i=1}^n \sum_{j=1}^m \frac{\beta_j q_{ij}^2}{2}  .
	\end{alignedat} \label{opt:NE}
	\end{equation} 
\end{lemma}

In general, the supply profile at the unique Nash equilibrium 
may not coincide with the  efficient supply profile. The potential loss of efficiency is can be measured according to the \emph{price of anarchy} of the underlying game \citep{KP99}. 
In the remainder of the paper, we use the terms \emph{worst-case efficiency loss} and \emph{price of anarchy} interchangeably. 
\begin{definition}[Price of Anarchy] \label{def:PoA} \emph{Given an edge set $\Ecal$ and cost function profile $C$, we define the \emph{price of anarchy (PoA)} associated with the corresponding networked Cournot game $(F, \mathcal{Q}(\Ecal), \profit)$  as
	\begin{align*}
	\rho (\Ecal, C) :=  \frac{\SW^* (F\times M, C)}{\SW \left( q^{\textrm NE} (\Ecal), C \right) }.
	\end{align*}
	 We define $\rho (\Ecal, C) := 1$ if both $\SW^* (F \times M, \cost)  = 0$ and $\SW (\production^{\rm NE} (\Ecal), \cost)=0$.} 
\end{definition}

It is worth noting that, in games that may exhibit a multiplicity of Nash equilibria, the price of anarchy is typically defined as the ratio of the efficient social welfare over that of the Nash equilibrium with the \emph{lowest} social welfare. 
Since each of the networked Cournot games considered in this paper admits a unique Nash equilibrium, \eb{this more general definition of price of anarchy reduces to the one specified in Definition \ref{def:PoA}.}
It's also worth mentioning that, in our definition of price of anarchy, the social welfare at Nash equilibrium is always measured against the efficient social welfare, which is defined  in terms of the complete edge set $F \times M$.


\subsubsection{Competition under controlled allocation platforms:}
We model competition in controlled allocation platforms according to a multi-leader, single-follower Stackelberg game. 
\eb{In this context,  firms lead by determining their aggregate production quantities simultaneously, in anticipation of the  platform's allocation of the resulting aggregate supply across the different markets according to \eqref{opt:cont}, and the subsequent determination of price according to \eqref{eq:capprice}.}
\eb{Letting $s_i \in \Rset_+$ denote the aggregate quantity supplied by each firm $i\in F$, the payoff derived by each firm under the controlled allocation \eqref{opt:cont}, pricing mechanism \eqref{eq:capprice}, the platform objective defined by $\lambda$, and its corresponding allocation function $A_\lambda$, can be explicitly defined as 
\begin{equation}
\pi_i(s_i, s_{-i}; \lambda) := s_i p\left(A_\lambda\left(\sum_{i=1}^n s_i\right)\right) -C_i(s_i).
\end{equation}}
\rone{We overload the use of the notation $\pi_i$ in the setting here in a slight abuse of notation, but the operating definition will be clear from the context. Based on this, we introduce the definition of a Stackelberg equilibrium in the networked Stackelberg game. }
\rone{\begin{definition}[Stackelberg Equilibrium]
\emph{The supply profile $s \in \Rset_+^n$ constitutes a \emph{Stackelberg equilibrium} of the networked Stackelberg game with platform objective parameterized by $\lambda$ if $\pi_i(s_i,s_{-i}; \lambda) \ \geq \ \pi_i(\overline{s}_i,s_{-i}; \lambda)$ for all $\overline{s}_i \in \Rset_+$ and  every firm $i \in F$. When multiple equilibria exists, we let $\rm{SE} \subset \Rset_+^n$ denote the \emph{set of Stackelberg equilibria} associated with the networked Stackelberg game.} 
\end{definition}}

\eb{We note that the family of  networked Stackelberg games studied in this paper is similar in spirit to the class of games  considered in \citet{cai2017role, xu2017efficiency}.}  In this earlier work, it is shown that networked Stackelberg games may possess (i) no Stackelberg equilibrium, (ii) a unique Stackelberg equilibrium, or (iii) multiple Stackelberg equilibria.
\eb{Taking these different possibilities into account, we define the price of anarchy associated with the networked Stackelberg game as follows. }
\rone{\begin{definition}
	\emph{The \emph{price of anarchy} $\rho(\lambda, M)$ associated with the networked Stackelberg game with platform objective function determined by $\lambda$ and a set of markets $M$ is defined as 
    \eb{$$
    \rho(\lambda, M) := \sup_{s \in \rm{SE}} \ \frac{\SW^*(F\times M,C)}{\SW(s, \lambda)},
    $$}
    with set of firms $F$ having the corresponding set of cost functions $C$ . We define $\rho(\lambda, M) := \infty$ if $\rm{SE} = \emptyset$, and $\rho (\lambda, M) := 1$ if both $\SW^* (F \times M, \cost)  = 0$ and $\SW(s, \lambda)$ = 0.}
\end{definition}
In this context, we have to chosen to make explicit the dependence of the price of anarchy on the platform parameter $\lambda$ and the set of markets $M$ to facilitate the presentation of our main results. We also note that, while we have reused the notation $\rho(\cdot, \cdot)$  in  defining  the  price of anarchy function associated with each platform design, the specific platform being considered will always be clear from the surrounding context.}
\rone{\subsection{Illustrative applications}} \label{sec:applications}

\rone{In what follows, we discuss connections of our platform models to ridesharing and e-commerce.  
Note that we are not seeking to model all details of these applications. Rather, our objective is to illustrate the relevance and limitations of applying our networked Cournot model to these practical settings. 
}

\rone{\emph{Ridesharing.} \sam{Ridesharing platforms can be modeled as a controlled allocation platform. Specifically, collection of drivers located in a common geographical region can be approximately modeled as a single firm,} \sam{where the number of rides provided by drivers in that regions can be modeled 
as the firm's production quantity.
Importantly, a supply-side model of this type can capture the strategic withholding of supply (rides) from the drivers.
A market, on the other hand, corresponds to a collection of customers whose current locations and ride destinations correspond to a specific pair of geographical regions.
In practice, the platform matches each ride request to an available driver nearby according to the solution of an online matching problem. In the context of controlled allocation platforms, we provide a simplified description of this matching mechanism according to the market clearing mechanism specified in \eqref{opt:cont}.}
} 

\sam{Modeling ridesharing platforms in this manner enables the analysis of the efficiency loss in ridesharing platforms that is associated with the supply withholding from the driver side. In Section \ref{sec:cap}, we show that such efficiency loss might be infinitely large for a range of market clearing mechanisms. The limitation of this model, however, is that it does not enable the description of the dynamic evolution of the supply and the demand side that is driven by the spatial movements of the drivers and the riders.}

\rone{\emph{E-commerce.}} \sam{Depending on its method of access control, e-commerce platforms like Ebay, Amazon, or Alibaba can be described as open access or discriminatory access platforms.}
\rone{For example, Amazon can be viewed as a discriminatory access platform, where the Buybox is used to identify and highlight a single seller based on performance measures such as pricing, reviews and shipping. In contrast, eBay and Etsy can be viewed as open access platforms. Alibaba serves similarly as a platform between manufacturers, suppliers/exporters, and importers. }
\sam{In this paper, we quantify the worst-case efficiency loss in open access and discriminatory access platforms. Our results reveal insights on the efficiency gain that can be attained by access control in e-commerce platforms.}

\sam{
Specifically, we use the networked Cournot model to describe the interaction between the sellers of a single product in e-commerce platforms.
A market corresponds to the aggregation of consumers in a particular geographical region.  A producer, on the other hand, corresponds to one seller of the single product under consideration. 
The set of markets that each producer has access to depends on the method of access control in the platform.
The decision variable of each producer is assumed to be his production quantity at each market he has access to.
Note that although producers do not directly specify their production quantities in e-commerce platforms, this Cournot model is still potentially useful in helping predict the production capacity that each firm builds out in the long run.
Specifically, as is pointed out in \citet{kreps1983quantity}, the price and production levels at the unique Cournot equilibrium are identical to the price and production levels at the unique subgame perfect equilibrium of a two-stage game, in which firms first choose their capacities (\`a la Cournot) in Stage 1, followed by a choice of prices (\`a la Bertrand) in Stage 2, where production levels are set according to the capacity levels chosen in Stage 1.
}

%% file: openaccess.tex

Open access platforms, such as eBay and Etsy, provide all customers access to all active sellers.
Such platforms are typically accompanied by lower barriers to entry, which has the potential to increase market competitiveness and efficiency.
%
In this section, we  provide tight bounds on the worst case efficiency loss in  open access platforms under a variety of assumptions on supply and demand-side characteristics.


\begin{theorem} \label{thm:PoA_asym}
	The worst case efficiency loss associated with a cost function profile $C$ and the corresponding open access networked Cournot game $(F, \mathcal{Q}(F \times M), \pi)$ is upper bounded by
	\begin{align*}
	\rho (F \times M,C) \leq \frac{3}{2} \left( 1 - \frac{1}{3n+6} \right).
	\end{align*}
 	The bound is tight if all markets are identical in terms of their maximal willingness to pay. That is, if  $\alpha_1 = \alpha_2 =  \cdots = \alpha_m$, then there exists a cost function profile $\overline{C}$ such that
\begin{align*}
	\rho (F \times M,\overline{C}) =  \frac{3}{2} \left( 1 - \frac{1}{3n+6} \right).
\end{align*}
\end{theorem}

\noindent \eb{Letting the number of firms $n \rightarrow \infty$, we have the following result as an immediate corollary to Theorem \ref{thm:PoA_asym}. }

\begin{corollary}
	\label{cor:poabound} The price of anarchy of open access platforms is at most 3/2.
\end{corollary}
\eb{The price of anarchy bound in \autoref{cor:poabound} is valid for open access platform with any number of firms and markets. It, therefore, recovers the 3/2 price of anarchy bound first established by \citet{johari2005efficiency} in the context of single-market Cournot games as a special case.}

\eb{In addition to having bounded efficiency loss,  open access platforms are also guaranteed to limit the extent to which firms withhold their production (relative to socially optimal levels) at  Nash equilibrium.} \sam{We formally state this property in Proposition \ref{lem:prodpres}.}

\begin{proposition}
    \label{lem:prodpres}
    \rone{
	  At each Nash \eb{equilibrium}, the demand fulfilled in each  market $j\in M$ is at least half of the demand fulfilled at the socially efficient production profile.} 
\end{proposition}

\sam{In other words, the amount of production that is withheld in each market at Nash equilibrium is no greater than half of the socially efficient production quantity.  This ensures that aggregate consumer utility at Nash equilibrium is at least $3/4$ of the aggregate consumer utility at the efficient production profile---a result that is critical to the low efficiency loss of open access platforms. }

\subsection{Excess Entry in Open Access Platforms }
\eb{Theorem \ref{thm:PoA_asym} reveals that increased competition in the supply-side of the platform does not necessarily translate to increased market efficiency.
We shed  light on what drives this seemingly counterintuitive behavior, where the entry of a ``new'' firm to the platform can result in a reduction of social welfare at Nash equilibrium. 
While the entry of a new firm is guaranteed to increase the aggregate quantity supplied to each market at equilibrium, the entry of an additional firm will also result in a reduction in the  quantities produced by each of the existing firms at equilibrium. Thus, if the new entrant's marginal cost to produce is large relative to existing firms, then its entry will result in an increase in the aggregate production cost at equilibrium. And, under certain conditions on the firms' cost structure, this increase in the aggregate production cost will exceed the corresponding increase in aggregate consumer utility driven by the new entrant---resulting in an overall reduction in social welfare at equilibrium.} \sam{This behavior is closely related to the  ``excess entry theorem'' in the economics literature  \citep{suzumura1987entry, mankiw1986free, lahiri1988helping}.
}

We investigate this behavior more deeply in Sections \ref{sec:worst_case_cost}-\ref{sec:efflossOAP}, where we provide a refinement of the price of anarchy bound in Theorem \ref{thm:PoA_asym} that sheds light on the role played by asymmetry in the cost structure across firms. As part of these analyses, we  also establish a technical lemma in Section \ref{sec:worst_case_cost}, which characterizes the functional form of the cost function profile that maximizes the efficiency loss in networked Cournot games.

\subsection{Worst-Case Cost  Profiles} \label{sec:worst_case_cost}

\eb{The following result reveals that, among all convex production cost functions satisfying Assumption \ref{ass:cost}, \emph{piecewise linear cost functions maximize the worst-case efficiency loss in networked Cournot games.} The result is also constructive in nature. That is, given a particular game and  cost profile satisfying Assumption \ref{ass:cost}, Lemma \ref{Lem:WorstCaseLinear} provides an explicit characterization of a piecewise linear cost profile that gives rise to a new game whose price of anarchy is no less than that of the original game. Thus, in constructing bounds on the price of anarchy associated with a particular game, it suffices to consider piecewise linear cost functions of the form specified in Lemma \ref{Lem:WorstCaseLinear}.}





\begin{lemma} \label{Lem:WorstCaseLinear}
	Let  $C = (C_1, \dots, C_n)$ be a cost function profile satisfying Assumption \ref{ass:cost}. Let $q^{\rm NE}( F \times M)$ denote the unique Nash equilibrium of the corresponding  networked Cournot game, and define the piecewise linear cost function profile $\overline{C} = (\overline{C}_1, \dots, \overline{C}_n)$ according to
	$$\overline{C}_i (s_i) := \left( \partial^+ C_i \left( \sum_{j=1}^m q_{ij}^{\rm NE} ( F \times M ) \right) \cdot s_i \right)^+$$
	for $i=1,\dots,n$, where $\partial^+ C_i$ denotes the right-derivative of the function $C_i$. It holds that  $$\rho (F \times M , C) \, \leq \,  \rho (F \times M, \overline{C} ).$$
\end{lemma}
We provide a complete proof of Lemma \ref{Lem:WorstCaseLinear} in the Appendix to this paper. The characterization of the worst-case cost  profile in Lemma \ref{Lem:WorstCaseLinear} will prove crucial to the derivation of tight upper bounds on the price of anarchy for the class of networked Cournot games considered in this paper. \eb{Specifically, 
under linear cost profiles, a networked Cournot game can be decoupled into $m$ single-market Cournot games,  each  corresponding to a distinct market in the original networked game.}



\subsection{Efficiency Loss in Open Access Platforms}

\label{sec:efflossOAP}
\eb{Theorem \ref{thm:PoA_asym} reveals the possibility that increased competition in open access platforms may not always lead to increases in  platform efficiency. In what follows, we this investigate this behavior in more depth by examining the impact that producer  \emph{cost asymmetry}  has on  platform efficiency.}

\subsubsection{Symmetric Cost Profiles}\label{subsec:symm}
We first consider the symmetric setting, where firms are assumed to have identical cost functions. 
\eb{In Proposition \ref{prop:PoA_symm}, we establish a tight upper bound on the price of anarchy in this setting. The upper bound that we construct \emph{decreases monotonically} in the number of firms, and converges to one as the number of firms grows large---a behavior that is consistent with  classical competitive limit theorems for markets with symmetric firms \citep{mas1995microeconomic}.}



\begin{proposition} \label{prop:PoA_symm}
	Assume that $C_1 = C_2 = \cdots = C_n$.  The price of anarchy associated with the corresponding networked Cournot game $(F, \mathcal{Q}(F \times M), \pi)$ is upper bounded by
	\begin{align*}
	\rho (F \times M,C) \, \leq \, 1 + \frac{1}{(n  + 1)^2 - 1}.
	\end{align*}
	Additionally, this bound is tight. That is, there exists a symmetric cost  profile under which the upper bound is achieved.
\end{proposition}


We apply  \autoref{Lem:WorstCaseLinear} in proving Proposition \ref{prop:PoA_symm}, which reveals that the worst-case symmetric cost  profile is comprised  of $n$ identical cost functions that are linear on $(0, \infty)$. The complete proof of  \autoref{Lem:WorstCaseLinear} can be found in the  Appendix.

\subsubsection{Asymmetric Linear Cost Profiles}\label{subsec:asymm}
\eb{At first glance, the efficiency loss bounds established  in Proposition \ref{prop:PoA_symm} and Theorem \ref{thm:PoA_asym} may appear to stand in conflict.
If firms are symmetric, then the efficiency loss bound that we construct \emph{decreases} monotonically in the number of firms, approaching one asymptotically. If there is asymmetry in the cost structure across firms, however, then the corresponding efficiency loss bound that we establish \emph{increases} monotonically in the number of firms, approaching 3/2 as the number of firms grows large.
In what follows, we resolve this seeming discrepancy  by refining the structure of our efficiency loss bounds to reflect an explicit measure of cost asymmetry between firms. To facilitate our analysis, we restrict our attention to cost functions} that are linear on $(0, \infty)$, and whose slopes lie within  $[c_{\min}, c_{\max}] \subseteq \mathbb{R}_+$. We denote the space of all such cost functions by
\begin{align*}
\mathcal{L} (c_{\min}, c_{\max}) := \left\{C_0: \Rset \to \Rset_+ \;  \Big| \; C_0(x) = \left( cx\right)^+, \right. 
~  c \in [c_{\min}, c_{\max}] \Big\}&.
\end{align*}
We write $C \in \mathcal{L}^n (c_{\min}, c_{\max})$ if the cost function profile $C$ satisfies $C_i \in \mathcal{L} (c_{\min}, c_{\max})$ for each  firm $i \in F$. 
\eb{We define a non-dimensional measure of cost asymmetry that is normalized according the maximum willingness to pay in each  market $j \in M$ according to}
\begin{align}
\gamma_j :=  1 - \frac{ c_{\max} - c_{\min}}{\alpha_j - c_{\min}}.
\end{align}
It holds that $\gamma_j \in (-\infty, 1]$ if $c_{\min} < \alpha_j$. 
Clearly, $\gamma_j$ is increasing in consumers' maximum willingness to pay $\alpha_j$, and decreasing in the maximum marginal cost $c_{\max}$.  A value of $\gamma_j$ close to one implies a small degree of asymmetry between firms' cost functions relative to consumers' maximum willingness to pay in market $j$.



The following result provides a tight bound on the price of anarchy in open access platforms when firms have linear cost functions with a bounded degree of asymmetry. 
\begin{proposition}
	\label{prop:linearcostspoa} Let $C \in \mathcal{L}^n (c_{\min},  c_{\max})$, and assume that 
	$c_{\min} < \max_{j\in M} \,  \{\alpha_j\}$. 
	The price of anarchy of  the corresponding networked Cournot game $(F, \mathcal{Q}(F \times M), \pi)$  satisfies
	\begin{align*}
	\rho (F \times M , C) \leq \dfrac{\displaystyle\sum_{j=1}^m  \frac{\left( (\alpha_j - c_{\min} )^+ \right) ^2}{\beta_j}}{\displaystyle\sum_{j=1}^m \left( \frac{2n+4}{3n+5}  + \delta (\gamma_j, n) \right) \frac{ \left( (\alpha_j - c_{\min} )^+ \right)^2}{\beta_j}},
	\end{align*}
	where 
		\begin{align*}
	\delta (\gamma, n): = \begin{cases}
		\displaystyle\frac{(n-1) (3n+5)}{(n+1)^2} \left(  \gamma - \frac{2n+3}{3n+5} \right)^2, & \text{if } \displaystyle\gamma \geq \frac{2n+3}{3n+5}, \\
0,  & \text{otherwise } .
	\end{cases}
	\end{align*}
	The bound is tight if $\alpha_1 = \alpha_2 = \cdots = \alpha_m$.
	\label{prop:PoA_cmin_cmax}
\end{proposition}


The price of anarchy bound established in Proposition \ref{prop:PoA_cmin_cmax} captures the impact that cost asymmetry has on efficiency loss through the 
market-dependent quantities $\delta (\gamma_j, n)$  for $j =1, \dots, m$. Crucially, each term  $\delta (\gamma_i, n)$ is nondecreasing in $\gamma_j$, revealing that a reduction in the degree of asymmetry in the cost structure across  firms translates to a reduction in the corresponding price of anarchy bound. \eb{As a corollary to this result, the price of anarchy bound in Proposition \ref{prop:PoA_cmin_cmax} reduces that of Proposition \ref{prop:PoA_symm} in the symmetric limit where $\gamma_j =1$ for all markets $j \in M$.}

%% file: controlledplatform.tex
\sam{While open access platforms are shown to exhibit bounded efficiency loss, the welfare loss stemming  from  the  ``misallocation"  of  supply  across  the  different  markets  can  be  large.
In this section, we investigate a class of  \emph{controlled allocation} platform designs, where the platform optimally allocates the aggregate production across the different markets to maximize  a convex combination of consumer surplus and producer revenue.
Despite their welfare-maximizing intentions, controlled allocation platforms may suffer large efficiency loss in practice, due to the firms' withholding their production relative to socially optimal levels. Such  withholding of capacity is has been observed to occur in ride-sharing platforms, for example \citep{chen2017thrown}.
In this section, we show} \rone{that controlled allocation platforms may exhibit an unbounded price of anarchy. In particular, when the platform controls the allocation of supply to maximize social welfare, the corresponding price of anarchy is $\Omega(m)$,} \eb{which grows unbounded in the number of markets.  This is in stark contrast to open access and discriminatory access platforms, which are guaranteed to have a price of anarchy that is uniformly bounded from above in the number of markets $m$.}

\sam{We first consider a class of controlled allocation platform that maximize social welfare. This corresponds to a platform objective function ${\rm OBJ}(d;1/2)$ (i.e., $\lambda = 1/2$). In Theorem \ref{thm:ineff}, we show that the price of anarchy of controlled allocation platforms that maximize social welfare grows at least linearly in the number of markets.}
\begin{theorem}
	\label{thm:ineff}
	\sam{The price of anarchy of a controlled allocation platform that maximizes social welfare ${\rm OBJ}(d;1/2)$ grows at least linearly in the number of markets. That is, there exists a family of inverse demand functions and cost profiles such that}
    $$
    \rho(1/2, M) \geq \frac{8m}{9}.
    $$
\end{theorem}



\sam{
The primary driver behind this growth in efficiency loss  stems from the production withholding incentives  that emerge in controlled allocation platforms. In open access platforms, each firm is free to specify the exact quantity that it produces in each market. As a result, a firm can  increase its production in one market without affecting its revenue in other markets. 
In contrast, in controlled allocation platforms,  each firm is only allowed to specify its aggregate production, while the platform determines the allocation of supply across the different markets. 
Crucially, increasing a firm's aggregate production may result in a reduction in the uniform price that the firm is paid (in all markets), which may  ultimately give rise to stronger incentives for  production withholding than in open access markets.
In the proof of Theorem \ref{thm:ineff}, we construct an example that clearly illustrates this intuition. 
}

\eb{In Section \ref{sec:lb_cap}, we extend these analyses to address the entire gamut of controlled allocation  optimization objectives  ${\rm OBJ}(d;\lambda)$ ($0 \leq \lambda  \leq 1$) considered in this paper. Loosely speaking, our results reveal that platform objectives that ``favor'' firms over consumers  (by weighing the firms' aggregate revenue more heavily) weaken the incentive for production withholding.}

\subsection{Lower Bounds on the Price of Anarchy in Controlled Allocation Platforms} \label{sec:lb_cap}

 \sam{We first consider the case in which the  platform controls the allocation of supply to maximize  consumer surplus ($\lambda = 1$).}
 
 




\begin{proposition}\label{claim:cap-cs}
	Assume that there are at least $m \geq 2$ markets.
	Controlled allocation platforms that maximize  consumer surplus have an unbounded price of anarchy. \sam{That is, there exists a family of inverse demand functions and cost profile such that}
	\begin{align*}
	    \rho (1, M) = \infty.
	\end{align*}
\end{proposition}

\sam{
It is not difficult to show that controlled allocation platforms, which maximize ${\rm OBJ}(d;1)$ (consumer surplus),  will allocate the entire aggregate supply to the market that has the largest demand price elasticity.
A supply allocation rule of this form will, in general, result in  production withholding relative to socially optimal levels. We illustrate these effects more concretely in a simple two-market platform.} In this example, each 
 firm is assumed to have a constant marginal cost of production $c > 0$, and the inverse demand functions associated with each of the two markets are defined as 
$$
p_1 = (c+\epsilon) - (\epsilon/2) d_1  \ \ \text{and} \ \ \ p_2 = (c-\epsilon) - \epsilon d_2, 
$$
where $\epsilon > 0$.
\sam{The supply allocation rule, which maximizes consumer surplus, will allocate the aggregate production quantity supplied by the firms to the second market. As the resulting market clearing price will in general be smaller than the per-unit production cost of each firm, the firms will collectively withhold their output, producing nothing at Nash equilibrium.} \eb{This is in contrast to  the socially efficient market outcome, where firms collectively produce two units in the first market and zero units in the second market. The price of anarchy is, therefore, unbounded in this simple example. }

\sam{Proposition \ref{claim:cap-cs} reveals that a poorly chosen market clearing mechanism may result in a substantial loss of efficiency at Nash equilibrium.  In what follows, we inspect a family of market clearing mechanisms where the platform's  objective function ${\rm OBJ}(d;\lambda)$ ($0 \leq \lambda  \leq 1$) is specified according to a convex combination of aggregate consumer surplus and producer revenue (cf. Section \ref{sec:prelims} for detailed formulation).
%
In Theorem \ref{thm:generalcap}, we provide lower bounds on the price of anarchy in the controlled allocation platforms as a function of the objective function being optimized by the platform. }
\rone{\begin{theorem}\label{thm:generalcap}
    \sam{Assume that there are at least $m \geq 2$ markets.} The price of anarchy $\rho(\lambda,M)$  in  controlled allocation platforms satisfies: 
\begin{equation*} 
\rho(\lambda, M) \geq 
\begin{cases}
\max \left\{ \displaystyle\frac{3}{2}, \  \frac{2}{3}\left(1+\sqrt{1+\frac{\lambda^2}{(2 \lambda -1)(\lambda -1)}}\right)  \right\},  &  \textrm{if} \ \  \lambda \in [0, \, 1/2)\\
\displaystyle\frac{8m}{9}, & \textrm{if} \ \ \lambda \in [1/2, \, 2/3) \\
\infty, & \textrm{if} \ \ \lambda \in [2/3, \, 1]\\
\end{cases}
\end{equation*}
\end{theorem}}

\sam{
If we let the number of markets $m \to \infty$, then the lower bound on the price of anarchy in controlled allocation in Theorem \ref{thm:generalcap} increases from $3/2$ to $+\infty$ as $\lambda$ increases from 0 to $1/2$, and remains $+ \infty$ for $\lambda \in [1/2, 1]$. Such a trend in the price of anarchy bound shows that the efficiency loss stemming from the incentive for production withholding potentially decreases as we increase the extent to which the platform's objective favors firms.
}



The results contained in this section reveal that the  explicit control of the allocation of supply to each market by the platform can result in  production withholding by firms and large efficiency loss at Nash equilibrium. \eb{In the following section, we explore an alternative approach to platform control, where the platform employs a moderate approach to control by restricting the set of markets that different firms are given access to, while allowing firms to freely determine their individual production levels in each market. We refer to such platforms as \emph{discriminatory access platforms.}}

%% file: discplatform.tex
\eb{In this section, we investigate a class of  \emph{discriminatory access platforms} that aim to strike a balance between open access and controlled allocation platform designs, in order to improve upon on the worst case efficiency loss of both. As a compromise between these two extremes, discriminatory access platforms exercise a form of limited control in the marketplace by restricting the set of markets that different firms are given access to, while allowing firms to freely determine their individual production levels in each market.  Platforms designs of this kind are reminiscent of Amazon's \emph{Buybox} and Airbnb's \emph{Superhost} mechanisms, which selectively display particular suppliers to customers, while hiding others.}



In the context of our model, this form of access control  corresponds to the specification  of the edge set $\Ecal \subseteq F \times M$ in the bipartite graph that determines connections between firms and markets, with the goal of maximizing social welfare at the unique Nash equilibrium of the resulting networked Cournot game (cf. problem \eqref{opt:network}). 
Importantly, unlike the controlled allocation platforms, the efficiency loss of discriminatory access platforms is guaranteed to be no worse than that of open access platforms, since open access designs represent a particular approach to access control.

\eb{Utilizing an assumption of \emph{linearity} of the firms' cost functions, in Theorem \ref{thm:discriminatory_PoA}, we provide  an explicit bound on the price of anarchy  in discriminatory access platforms} under the optimal network design.
 We build on this result in Theorem \ref{thm:greedy-alg}, showing that the optimal network design can be  computed using an intuitive greedy algorithm that we specify later in the section. 

\begin{theorem} \label{thm:discriminatory_PoA}
	Let $C \in \Lcal^n (c_{\min}, c_{\max})$  and assume that $c_{\min} < \max_{j\in M} \alpha_j$.  Additionally, let $\Ecal^* \subseteq F \times M$ denote the network structure that maximizes the  social welfare derived  under discriminatory access platforms at Nash equilibrium according to problem \eqref{opt:network}. The efficient social welfare associated with the edge set $\Ecal^*$ satisfies
	$$\SW^* (\Ecal^*, C) = \SW^* (F \times M, C).$$
	Moreover, the price of anarchy of the corresponding networked Cournot game $(F, \mathcal{Q}(\Ecal^*), \pi)$ is upper bounded by
	\begin{align*}
	\rho (\Ecal^*, C) 
	\leq \dfrac{\displaystyle\sum_{j=1}^m  \frac{\left( (\alpha_j - c_{\min} )^+ \right)^2}{\beta_j}}{\displaystyle\sum_{j=1}^m \underset{k \in \{1, .., n\}}{\max}\left\{ \frac{2k+4}{3k+5}  + \delta (\gamma_j, k) \right\} \frac{ \left( (\alpha_j - c_{\min} )^+ \right)^2}{\beta_j}}.
	\end{align*}
	Recall that the function $\delta(\cdot,\cdot)$ (originally defined in Proposition \ref{prop:PoA_cmin_cmax})  measures the degree of asymmetry in the firms' cost profile. Finally, the above bound on the price of anarchy is tight if $\alpha_1 = \alpha_2 = \cdots = \alpha_m$.
\end{theorem} 
\eb{It is important to note that, while we have restricted ourselves to linear cost profiles in Theorem \ref{thm:discriminatory_PoA}, the upper bound on the price of anarchy that we have derived also holds in the more general setting where firms' have convex cost functions. This follows from Lemma \ref{Lem:WorstCaseLinear}, which shows that, among all convex cost profiles,  linear cost profiles result in the greatest loss of efficiency at equilibrium.}
\eb{Additionally, the upper bound on the price of anarchy that we provide is maximized at $n=1$, resulting in 
a price of anarchy bound} of $4/3$ that holds for \emph{optimized} discriminatory access platforms with any number of firms and markets. This improves substantially upon the $3/2$ upper bound on the  price of anarchy in open access platforms (cf. Corollary \ref{cor:poabound}). The result is formally stated as follows. 
\begin{corollary}
	\label{cor:nine}
	The price of anarchy of optimized discriminatory access platforms is at most $4/3$. 
\end{corollary}



\eb{In contrast to controlled allocation platforms, which suffer from an unbounded price of anarchy, the price of anarchy of optimized discriminatory access platforms is not only guaranteed to be bounded, but also improves upon the  worst case efficiency loss  of open access platforms.}


\subsection{The Optimal Network Design Problem}

The optimal network design problem amounts to the selection of an 
edge set $\Ecal$ that maximizes social welfare at the unique Nash equilibrium of the resulting networked Cournot game. Formally, Lemma \ref{lem:NE} provides a characterization of the supply profile at the unique Nash equilibrium of the game $(F, \Qcal (\Ecal), \pi)$ as the unique optimal solution to a convex program. Therefore, the \emph{optimal network design problem} can be formulated  as the  following mathematical program with equilibrium constraints (MPEC):
\begin{equation} \label{opt:network}
\begin{alignedat}{8}
& \text{maximize} \quad && \SW (q, C) \\
& \text{subject to} \quad && \Ecal \subseteq F\times M \\
&&& q \in \underset{x \in \Qcal (\Ecal) }{\argmax} \left\{ \SW (x, C) - \sum_{j=1}^m  \sum_{i=1}^n \frac{\beta_j x_{ij}^2}{2}  \right\}
\end{alignedat}
\end{equation}
Here, the decision variables are the edge set $\Ecal$ and the supply profile $q$. 
The challenge in solving the above problem the presence of the discrete decision variables $\Ecal$, and the equilibrium constraint on $q$. An equilibrium constraint requires that a decision variable be an optimal solution to a optimization problem. In general, this can result in o a nonconvex and disconnected feasible region. See \citet{luo1996mathematical} for a more detailed discussion.
In what follows, we show that, when considering linear cost profiles, the proposed MPEC \eqref{opt:network}  can be efficiently solved using a greedy algorithm.







\subsection{Greedy Algorithm for Optimal Worst-case Network Design}
In this section, we restrict ourselves to cost functions that are linear over $(0, \infty)$.
Specifically, we assume that the cost function of each firm $i \in F$ satisfies
$C_i (s_i) = (c_i s_i)^+$,  where $c_i \geq 0$. 
Leveraging on the assumption of linearity, we propose a greedy algorithm  that is shown to solve the optimal network design problem \eqref{opt:network}. At a high level, the proposed greedy algorithm functions as follows.
For each market $j \in M$, the greedy algorithm visits firms in ascending order of marginal cost, and provides each firm that it visits access to market $j$ if its inclusion in that market increases social welfare at equilibrium.

\begin{algorithm}
	\caption{The Greedy Algorithm}\label{alg:linearDis}
	\begin{algorithmic}[1]
		\Require $c_1 \leq \cdots \leq c_n$.      
		\State Initialize edge set $\Ecal \gets \emptyset$.
		
		\For{$j=1$ to $m$}
		\State Initialize firm index $i \gets 1$.
		\State Initialize edge set $\widetilde{\Ecal} \gets \Ecal$.
		\Repeat
		\State Update edge set $\Ecal \gets \widetilde{\Ecal}$.
		\If{$i \leq n$}
		\State Set edge set $\widetilde{\Ecal} \gets \Ecal \cup (i, j)$.
		\State Set firm index $i \gets i + 1$.
		\EndIf
		\Until{$\SW (q^{\rm NE} (\widetilde{\Ecal} ), C) \leq \SW (q^{\rm NE} (\Ecal), C)$}.
		\EndFor
		
		\State 
		\Return $\Ecal$.
	\end{algorithmic}
\end{algorithm}





It is not difficult to see that Algorithm \ref{alg:linearDis} yields an edge set $\Ecal^* \subseteq F \times M$ that induces a  Nash equilibrium whose social welfare  is no smaller than that of the open access platform. 

\begin{theorem}\label{thm:greedy-alg}
Let $\Ecal^*$ be an edge set obtained from Algorithm \ref{alg:linearDis}. It follows that   $(\Ecal^*,q^{\textrm NE}(\Ecal^*))$ is an optimal solution to the MPEC $(\ref{opt:network})$.
\end{theorem}


While Theorem \ref{thm:discriminatory_PoA} reveals a key advantage that optimized discriminatory access platforms have over open access platforms in terms of limiting efficiency loss at equilibrium, Theorem \ref{thm:greedy-alg} provides an algorithmic approach to efficiently  to  compute the optimal network structure in discriminatory access platforms.

%% file: conclusion.tex



\sam{
In this paper, we study the trade-off between transparency and control that arises in the design of access and allocation control mechanisms in online platforms.
Specifically, we leverage on a networked Cournot model to study the efficiency loss under three different platform designs: open access, controlled allocation and discriminatory access platform designs. 
}
%
%

Open access platforms, exemplified by eBay, allow all firms to participate freely in all markets.
\sam{Under the networked Cournot model considered in this paper, we show that open access platforms are guaranteed to preserve at least $2/3$ of the efficient social welfare at Nash equilibrium. In spite of this, the allocation of aggregate production among producers and markets is, in general, suboptimal in open access platforms. Specifically, our results reveal an ``excess entry theorem" in open access platforms, in the sense that the entry of new producers might lead to both a reduction in social welfare at Nash equilibrium if producers’ cost functions are allowed to be highly asymmetric.}


\sam{
In order to reduce the efficiency loss in open access control platforms that is associated with the potential ``misallocation" of supply between markets, we consider controlled allocation platforms, where the platform specifies the exact allocation of aggregate supply to each market. In spite of the good intentions behind the design of such platforms, the worst-case efficiency loss in controlled allocation platforms is unbounded due to the incentive for production withholding.
}

\sam{Finally, with the objective of striking a balance between open access and controlled allocation platform designs, we consider discriminatory access platforms, where the platform chooses the network that specifies the set of markets that each firm has access to.
We show that such a platform design improves upon open access platforms in terms of the worst-case efficiency loss.
Specifically, under the optimized network, discriminatory access platforms are guaranteed to preserve at least $3/4$ of the efficient social welfare. }


\sam{We conclude this paper with a discussion on future work.} One aspect of platform design that we have ignored thus far is related to customer \emph{search costs}. 
One of the initial goals of platforms is to lower search costs for consumers and lower entry costs for producers. Without careful understanding of the trade-offs between the aforementioned objectives and carefully balancing them, one can end up in a situation where search costs overwhelm consumers. For example, an open access platform design for a ride-sharing platform may not be a good idea, since the number of producers (in this case, drivers) in such platforms is large, resulting in large search costs for consumers. On the other hand, successful ride-sharing platforms, e.g., Uber, show only prices and do not offer alternative producers, eliminating any search cost involved in a transaction. Our preliminary results, included as Theorem 6 in the Appendix, confirm these observations when considering an additional search costs model. 
Additionally, certain unmodelled constraints can compel particular platform design over others. For example, electricity markets are an important platform that work under physical network constraints and therefore compelled to be under a controlled allocation design. As such, more work is needed to understand some of these issues, e.g., physical network constraints and the potential of demand management mechanisms such as demand response.

%% file: appendix.tex
\section{Proofs from Section \ref{sec:oap}}

\subsection{Proof of Lemma \ref{Lem:WorstCaseLinear}}
Let $q^\textrm{NE}(F \times M)\in \mathcal{Q}(F \times M)$ be the unique Nash equilibrium of the game $(F,\mathcal{Q}(F \times M),\pi)$ associated with an arbitrary cost function profile $\cost$. 
Throughout the proof, we always consider a networked Cournot game associated with the edge set $F \times M$. Thus, for notational simplicity we use $q^\textrm{NE}$ instead of $q^\textrm{NE}(F \times M)$ for the remainder of the proof.
For each $i\in F$, we define the non-negative scalar $\lambda_i$ according to
\begin{align}
\lambda_i := \partial^+ C_i \left( \sum_{j=1}^mq^\textrm{NE}_{ij} \right).
\end{align}

Here, $\lambda_i$ is the marginal cost of firm $i$ at the unique Nash equilibrium of the game $(F, \Qcal (F \times M), \pi)$. We define a (piecewise) linear cost function profile $\overline{\cost} = (\overline{\cost}_1, \dots, \overline{\cost}_n)$ according to
$$\overline{C}_i (s_i) := \max \{ \lambda_i s_i, \, 0\}, \quad i = 1, \dots, n.$$
Clearly, $\overline{C}_i$ is convex, and differentiable on $(0, \infty)$ for each $i \in \{1, \dots, n \}$. 
Additionally, the stationarity conditions of firms' profit maximization problem under the cost function profiles $C$ and $\overline{C}$ are identical at $q^{\textrm{NE}}$. The combination of these two facts shows that $q^{\textrm{NE}}$ is the unique Nash equilibrium of the game $(F,\mathcal{Q}(F \times M),\overline{\pi})$ associated with the cost function profile $\overline{C}$.
Our objective is to show that $\rho (F \times M, C) \leq \rho (F \times M, \overline{C})$, i.e.,
\begin{align*}
\frac{\SW (q^\textrm{NE}, C)}{\SW^*(F \times M,C)} \geq \frac{\SW (q^\textrm{NE}, \overline{C} )}{\SW^*(F \times M,\overline{C})}.
\end{align*}

In showing this, we first define the scalar $\mu_i$ for each firm $i\in F$ according to
\begin{align*}
\mu_i  :=  \partial^+ C_i \left( \sum_{j=1}^m q^\textrm{NE}_{ij}  \right) \cdot \left( \sum_{j=1}^m q^\textrm{NE}_{ij}  \right) - C_i \left( \sum_{j=1}^m q^\textrm{NE}_{ij}  \right).
\end{align*}
For each firm $i$, $\mu_i$ equals the absolute difference in his production cost at Nash equilibrium associated with the cost function profiles $C$ and $\overline{C}$.

With the definition of $\mu_i$ in hand, we define an ``intermediate'' cost function profile $\widetilde{\cost}= (\widetilde{\cost}_1, \dots, \widetilde{\cost}_n )$  according to:
$$\widetilde{C}_i (s_i) := \max \{ \lambda_i s_i - \mu_i, \, 0 \}, \quad i = 1, \dots, n.$$
The cost function profile $\widetilde{C}$ makes the connection between the cost function profiles $C$ and $\overline{C}$.
On one hand, for each firm $i \in F$, the cost function $\widetilde{C}_i$ can be regarded as a linearization of the original cost function $C_i$ around the Nash equilibrium of the game $(F, \Qcal(F \times M ), \pi)$. On the other hand, it can be interpreted as a translation of the cost function $\overline{\cost}_i$ downwards along the y-axis of length $\mu_i$, while keeping the resulting cost function non-negative for all real numbers. 
Note that the stationarity conditions of firms' profit maximization problem under the cost function profiles $C$ and $\widetilde{C}$ are identical at $q^{\textrm{NE}}$.
It follows that $q^\textrm{NE}$ is the unique Nash equilibrium of the game $(F, \Qcal (F \times M ), \widetilde{\pi} )$ associated with the cost function profile $\widetilde C$.

Since firms' production costs at $q^\textrm{NE}$ are equal for the cost function profiles $C$ and $\widetilde{C}$, we have that $\SW(q^\textrm{NE}, C) = \SW (q^\textrm{NE}, \widetilde{C})$. 
Moreover, since $C_i (s_i) \geq \widetilde{C}_i (s_i)$ for all $s_i \in \Rset_+$, we have that $\SW^*(F \times M,C) \leq \SW^*(F \times M,\widetilde{C} )$. It follows that
\begin{align}\label{eq:linearization}
\frac{\SW (q^\textrm{NE}, C)}{\SW^* (F \times M,C)} \geq \frac{\SW (q^\textrm{NE}, \widetilde{C} )}{\SW^* (F \times M, \widetilde{C})}. 
\end{align}

Additionally, the previous translation from $\overline{C}$ to $\widetilde C$ implies that $\SW (q^\textrm{NE}, \widetilde{C} ) $ and $\SW (q^\textrm{NE}, \overline{C})$ are related according to:
\begin{align}
\SW (q^\textrm{NE}, \overline{C}) = \SW (q^\textrm{NE}, \widetilde{C} ) - \sum_{i=1}^n \mu_i \geq 0. \label{eq:shifting}
\end{align}
We claim that the following inequality holds for the efficient social welfare $\SW^*(F \times M,\overline{C})$ associated with the cost function profile $\overline{C}$:
\begin{align*}
\SW^* (F \times M,\overline{C}) \geq \SW^* (F \times M,\widetilde{C}) - \sum_{i=1}^n \mu_i.
\end{align*}
To see this, let $q^*$ be an efficient supply profile under the cost function profile $\widetilde{C}$. For each $i\in F$, we have
\begin{align*}
\overline{C}_i \left( \sum_{j=1}^m q^*_{ij} \right) \leq \widetilde{C}_i \left( \sum_{j=1}^m q^*_{ij} \right) + \mu_i. 
\end{align*}
This inequality implies that
\begin{align*}
\SW^*(F \times M,\overline{C} ) &\geq \SW (q^*, \overline{C}) \geq \SW (q^*, \widetilde{C} ) - \sum_{i=1}^n \mu_i\\ 
&= \SW^* (F \times M,\widetilde{C} ) - \sum_{i=1}^n \mu_i,
\end{align*}
The above inequality, in combination with inequalities Equations \ref{eq:linearization} and \ref{eq:shifting}, shows that
\begin{align*}
&\frac{\SW (q^\textrm{NE}, C)}{\SW^* (F \times M,C)} \geq \frac{\SW (q^\textrm{NE}, \widetilde{C} )}{\SW^* (F \times M,\widetilde{C})} \\
&\qquad \geq \frac{\SW (q^\textrm{NE}, \widetilde{C} ) - \sum_{i=1}^n \mu_i}{\SW^* (F \times M,\widetilde{C}) - \sum_{i=1}^n \mu_i} \geq \frac{\SW (q^\textrm{NE}, \overline{C} )}{\SW^* (F \times M,\overline{C})},
\end{align*}
as needed to be shown.

\subsection{Proof of Proposition \ref{prop:PoA_symm}}
\noindent It follows from Lemma \ref{Lem:WorstCaseLinear} that the worst symmetric cost function profile that maximizes $\rho (F\times M, C)$ consists of $n$ identical cost functions that are linear on $(0, \infty)$. 
Thus, to upper bound the PoA of the networked Cournot game, it suffices to consider a cost function profile $\overline{C}$ that satisfies
\begin{align}
\overline{C}_i (x) = c x \quad \text{for } x \geq 0, \quad \text{for all } i \in F, \label{eq:cost_sym_linear}
\end{align} 
for a finite positive constant $c > 0$ that is independent of $i$.

Given the assumption on the (piecewise) linearity of cost, it is straightforward to show that the unique Nash equilibrium and an efficient supply profile of the corresponding networked Cournot game are given by
\begin{align*}
q^{\textrm{NE}}_{ij} = \frac{(\alpha_j - c)^+}{\beta_j(n+1)}, \quad \text{and} \quad q_{ij}^* = \frac{(\alpha_j - c)^+}{\beta_jn},
\end{align*}
respectively, for each $i \in F$, $j \in M$.
It follows that the social welfare at the unique Nash equilibrium $q^\textrm{NE}$ of this Cournot game is given by
\begin{align*}
\SW^* (q^{\textrm{NE}}, \overline{C}) = \sum_{j=1}^m\frac{\left( (\alpha_j - c)^+ \right)^2}{2 \beta_j} \left( 1 - \frac{1}{(n+1)^2} \right).
\end{align*}
And the efficient social welfare is given by
\begin{align*}
\SW^* (F\times M, \overline{C}) = \sum_{j=1}^m\frac{\left( (\alpha_j - c)^+ \right)^2}{2 \beta_j}.
\end{align*}
Hence, the price of anarchy associated with the cost function profile $\overline{C}$ is given by
\begin{align*}
\rho (F\times M, \overline{C} ) = \begin{cases}
1 + \frac{1}{(n  + 1)^2 - 1} & \text{if } \max_{j \in M} \alpha_j > c,  \\
1 & \text{otherwise } 
\end{cases}.
\end{align*}
Choosing $c < \max_{j \in M} \alpha_j$ gives the worst-case cost function profile that maximizes the price of anarchy over symmetric cost function profiles. This completes the proof.

\subsection{Proof of Theorem \ref{thm:PoA_asym}}

It follows from Lemma \ref{Lem:WorstCaseLinear} that the worst cost function profile that maximizes $\rho (F\times M, C)$ consists of cost functions that are linear on $(0, \infty)$.
%
We thus assume that each firm's cost function satisfies $\overline{C}_i (s_i) = (c_i s_i)^+,$ for $i = 1, \dots, n,$
where we assume without loss of generality that $c_1 \leq \cdots \leq c_n$.

Note that, for the case in which $n = 1$, a direct application of Proposition \ref{prop:PoA_symm} provides a tight price of anarchy bound of $\rho (F\times M, C) \leq 4/3$. 
For the remainder of the proof, we restrict ourselves to  $n \geq 2$.



We first consider the simple setting where the number of markets $m = 1$.
That is, the set of markets $M = \{1\}$.
Without loss of generality, we assume that $c_1 < \alpha_1$. 
Since cost functions are linear, there exists an efficient supply profile $q^*$ that assigns all production to firm 1,
i.e., $q^*_{i1} = 0 \quad \text{for } i = 2, \dots, n.$ 
The supply from firm $1$ and the corresponding efficient social welfare are given by: 
\begin{align*}
q_{11}^* = \frac{\alpha_1 - c_1}{\beta_1}, \quad \text{and } \quad \SW^* (F \times \{1\}, \overline{C}) = \frac{(\alpha_1 - c_1)^2}{2 \beta_1}.
\end{align*}
Fixing $\alpha_1$, $\beta_1$, $c_1$, we can optimize over $c_2, \dots, c_n$ in order to minimize the social welfare at the unique Nash equilibrium of the Cournot game. 
Using similar arguments as in the proof of \citep[Theorem 12]{johari2005efficiency}, one can formulate this problem as a symmetric convex program over the production quantities at Nash equilibrium of the remaining firms.
The optimal value of $c_2,\dots,c_n$ is given by 
\begin{align*}
c_i^* = \alpha_1 - \frac{2n+3}{3n+5} (\alpha_1 - c_1), \quad \text{for } i = 2, \dots, n.
\end{align*}
And the production quantity of each producer is given by
\begin{align*}
q_{i1} = \begin{cases}
\frac{\alpha_1 - c_1}{\beta_1} \cdot \frac{n+3}{3n+5} & \text{if } i = 1. \\
\frac{\alpha_1 - c_1}{\beta_1} \cdot \frac{1}{3n+5}, & \text{if } i = 2, \dots, n.
\end{cases}
\end{align*}
Define the cost function profile $\overline{C}^* = (\overline{C}^*_1, \dots, \overline{C}^*_n)$ according to $\overline{C}^*_1 (s_1) = (c_1 s_1)^+$ and $\overline{C}^*_i (s_i) = (c_i^* s_i)^+$ for $i = 2, \dots, n$.
Thus, for the fixed parameters $\alpha_1, \beta_1, c_1$, the minimum social welfare at Nash equilibrium is given by
\begin{align*}
\SW (q^{\rm NE} (F \times \{1\} ), \overline{C}^* ) = \frac{(n+2) (\alpha_1 - c_1)^2}{(3n+5) \beta_1}.
\end{align*}
It follows that for any linear cost function profile $\overline{C}$, we have
\begin{align*}
\rho (F \times \{1\}, \overline{C} ) \leq \rho (F \times \{1\}, \overline{C}^* )  = \frac{3}{2} \left( 1 - \frac{1}{3n+6} \right).
\end{align*}
%

We now consider the slightly more complicated setting in which $m >1$. 
The efficient social welfare associated with a linear cost function profile $\overline{C}$ is given by
\begin{align*}
\SW^* (F\times M, \overline{C} ) = \sum_{j=1}^m  \frac{\left( (\alpha_j - c_1 )^+ \right)^2}{2\beta_j}.
\end{align*}
Given the linearity of firms' cost functions over $(0, \infty)$, the networked Cournot game decouples across markets. Thus, the social welfare at Nash equilibrium satisfies
\begin{align*}
&\SW (q^{\rm NE} (F\times M), \overline{C}) = \sum_{j=1}^m \SW (q^{\rm NE} (F \times \{j\}), \overline{C}) \\
& \qquad \geq \sum_{j=1}^m \frac{\left( (\alpha_j - c_1 )^+ \right)^2}{ 3 \beta_j \left( 1 - \frac{1}{3n+6} \right) } 
= \frac{\SW^* (F\times M, \overline{C} )}{\frac{3}{2} \left( 1 - \frac{1}{3n+6} \right) }.
\end{align*}
It follows that the price of anarchy $\rho (F\times M, \overline{C} )$ satisfies
\begin{align*}
\rho (F\times M, \overline{C} ) \leq \frac{3}{2} \left( 1 - \frac{1}{3n+6} \right).
\end{align*}
Additionally, it is straightforward to check that this price of anarchy bound is achieved by the cost function profile $\overline{C}^*$ if $\alpha_1 = \alpha_2 = \cdots = \alpha_m$. This completes the proof.

\subsection{Proof of Proposition \ref{prop:PoA_cmin_cmax}}

\noindent We only provide a proof of the price of anarchy bound, since the proof on its tightness is straightforward.
The key intuition of the proof is that, under a linear cost function profile, the networked Cournot game decouples across markets.
The proof proceeds in two parts. 


\vspace{.05in}

\noindent \emph{Part 1:} 
We provide a price of anarchy bound for the case in which the number of markets $m = 1$. 
Namely, for any cost function profile $C \in \Lcal^n (c_{\min}, c_{\max})$, we have
$$\rho (F \times \{j\}, C) \leq \frac{1}{\frac{2n+4}{3n+5} + \delta (\gamma_j, n)}, \quad j = 1, \dots, m.$$
We omit the proof of this price of anarchy bound, as it follows from similar arguments as in the proof of Theorem \ref{thm:PoA_asym}.


\vspace{.05in}

\noindent \emph{Part 2:} We provide a price of anarchy bound for the case in which the number of markets $m >1$.
%
%
Let the cost function of each firm $i \in F$ be given by
$$C_i (s_i) = (c_i s_i)^+, \quad \text{where } c_i \in [c_{\min}, c_{\max}].$$
Without loss of generality, we assume that $c_1 \leq \cdots \leq c_n$. Thus, the efficient social welfare associated with the edge set $F\times M$ and the cost function profile $C$ is given by
$$\SW^* (F\times M, C) = \sum_{j=1}^m  \frac{\left( (\alpha_j - c_1 )^+ \right)^2}{2\beta_j}.$$
Since firms' cost functions are linear on $(0, \infty)$, the networked Cournot game decouples across markets. It follows that the social welfare at the unique Nash equilibrium of the game $(F, \Qcal (F\times M), \pi)$ satisfies
\begin{align}
\SW (q^{\rm NE} (F\times M), C) = \sum_{j=1}^m \SW (q^{\rm NE} (F \times \{j\}), C). \label{eq:pf_cmin_cmax_0}
\end{align}
Additionally, the term $\SW (q^{\rm NE} (F \times \{j\}), C)$ satisfies
\begin{align}
&\SW (q^{\rm NE} (F \times \{j\}), C) = \frac{\SW^* (F \times \{j\}, C) }{\rho(F \times \{j\}, C)  }\label{eq:pf_cmin_cmax_1}\\
& \qquad \geq  \sum_{j=1}^m \frac{\left( (\alpha_j - c_1 )^+ \right)^2}{2\beta_j} \left( \frac{2n+4}{3n+5} + \delta (\gamma_j, n) \right). \label{eq:pf_cmin_cmax_2}
\end{align}
Here, equation \ref{eq:pf_cmin_cmax_1} follows from the definition of the price of anarchy, and inequality \ref{eq:pf_cmin_cmax_2} follows from the price of anarchy bound in Step 1. A combination of Eq. \eqref{eq:pf_cmin_cmax_0} and inequality \eqref{eq:pf_cmin_cmax_2}  provides the following lower bound on the reciprocal of the price of anarchy $\rho (F\times M, C)$:
\begin{align*}
\frac{1}{\rho (F\times M, C)} \geq \dfrac{\sum_{j=1}^m \left( \frac{2n+4}{3n+5}  + \delta(\gamma_j, n) \right) \frac{ \left( (\alpha_j - c_1 )^+ \right) ^2}{\beta_j}}{\sum_{j=1}^m  \frac{\left( (\alpha_j - c_1 )^+ \right)^2}{\beta_j}}.
\end{align*}
One can verify that the partial derivative of the right-hand-side (RHS) of the above inequality with respect to $c_1$ is non-negative for $c_1 \in [c_{\min}, c_{\max}]$. 
%
Hence, choosing $c_1 = c_{\min}$ minimizes the RHS of the above inequality. This completes the proof.







\section{Proofs from Section \ref{sec:cap}}


\subsection{Proof of Theorem \ref{thm:ineff}}

\rone{
Consider $m$ many markets with the following market parameters: 
$$
p_1(d_1) = 1 -d_1, \ \text{and} \ p_j(d_j) = \frac{\theta^{j-1}}{1+\theta}- \frac{\theta^{2j-2}}{1-\theta^2}d_j, ~\forall j\geq 2,
$$
with $\theta<1/2$ and one costless producer participating in the presence of a social welfare maximizing controlled allocation platform. One can show that the social welfare maximizing controlled allocation platform allocates from market $1$ until price reaches that of market $2$, allocates between markets $1$ and $2$ keeping prices constant, until it approaches market $3$'s maximal willingness to pay, and allocates between the three, keeping prices constant in the different markets. Thereafter, whenever a new market is active, the platform allocates between all active markets, keeping prices constant, until price approaches the next market's maximal willingness to pay. We say a market becomes \emph{active} when supply starts entering it.  For a set of active markets, before the next market becomes active, the marginal increase in supply will be allocated in proportion to $1 / \demrate_j$ (in order to keep the prices the same).  This fully describes the behavior of the platform. In this case, it is convenient to first provide the price function as follows: 
$$
\price(A_{1/2}(x)) = \max_{0 \leq k < m} (\theta^k - \theta^{2k} x).
$$
Given the price function, one can in fact determine the quantity allocated to each market easily. One can observe which markets are active from the amount of quantity $x$. If $x$ sits on the $j$-th linear piecewise segment of this curve, then $j$ many markets are active. For each of these $j$ segments, the quantity in each segment is shared by its set of active markets based on the proportion $1/\beta_l$ for market $l$. These markets are designed such that the profit or revenue function are piecewise quadratic curves $x\cdot \price(A_{1/2}(x))$, which has for the $k$-th segment an optimal point $(2^{\theta^k})^{-1}$, with corresponding profit/revenue $\left({{2^\theta}^k}\right)^{-1}(\theta^k - \theta^{2k}\cdot ({2^{\theta^k})}^{-1}) = 1/4$. 
}

\rone{Each of those optimal points correspond to an equilibrium. The social welfare at the first equilibrium point is $3/8$. On the other hand, the efficient social welfare is obtained when quantity is produced to all the markets. Equivalently, the social welfare then is the area under the curve $\price(A_{1/2}(x))$. It is not hard to verify that $\price(A_{1/2}(x))$ is the piecewise linear function that connects the following points:
\begin{align*}
    (0, 1), \left(\frac 1 {1 + \theta}, \frac \theta {1 + \theta}\right), \left(\frac 1 {\theta(1+\theta)}, \frac{\theta^2} {1 + \theta}\right), \cdots \left(\frac 1 {\theta^{m - 2} (1 + \theta)}, \frac {\theta^{m - 1}} {1 + \theta}\right),\left(\frac 1 {\theta^{m-1}}, 0\right),
\end{align*} 
and the trapezoid whose vertices are $(1/ ({\theta^{k-1} (1 + \theta)}), 0)$,
$(1/({\theta^{k-1} (1 + \theta)}), {\theta^k}/({1 + \theta}))$,
$(1/({\theta^{k} (1 +\theta)}), 0)$,
$(1/({\theta^{k} (1 + \theta)}), {\theta^{k+1}}/({1 + \theta}))$ has area
\begin{align*}
\frac 1 2 \left( \frac {1}{\theta^k (1 + \theta)} - \frac {1}{\theta^{k-1} (1+ \theta)} \right) \left(\frac {\theta^k}{1 + \theta} + \frac{\theta^{k+1}}{1 +\theta} \right) = \frac {1 - \theta}{1+\theta}.
\end{align*}
There are $m-2$ such trapezoids under $\price(\demand)$, and therefore the socially efficient welfare is $\Omega(m)$. Taking $\theta=1/2$, the efficient social welfare is $m/3$, while the social welfare at the first equilibrium is $3/8$, and therefore price of anarchy is at least $8m/9$. }

Before providing a proof of \autoref{thm:ineff} we first state and prove the key structural lemma we use in the proof. The lemma essentially states that any convex, piecewise linear inverse demand curve with finitely many segments can be realized by the same number of linear demand markets under controlled allocation platforms which maximize social welfare. This is because such a platform would allocate from the market with the largest willingness to pay, and always keeping price constant. 



\begin{lemma}
	\label{lem:agg-demand}
	For a firm with costless production, a set of linear demand markets under controlled allocation is equivalent to a single market with a convex, piecewise linear demand curve.  Conversely, any convex, decreasing, piecewise linear demand curve with finitely many linear segments can be realized by a set of linear demand markets under a social welfare maximizing controlled allocation platform.
\end{lemma}

\proof{Proof of Lemma \ref{lem:agg-demand}.}
The characterization of the socially efficient production, with $\supply$ fixed highlights that the platform will reallocate this amount to $\demand_1, \ldots, \demand_m$ such that $\sum_j \demand_j = \supply$, and for each market~$j$ where $\demand_{j} > 0$, $\price_j$ is equivalent to a fixed price~$\price$ across markets; for each market~$j$ where $\demand_j = 0$, it must be that $\demlim_j \leq \price$.  This shows that, as $\supply$ increases, the allocation will enter the markets one by one in the order in which $\demlim_j$ decreases.  

We say a market becomes \emph{active} when supply starts entering it.  For a set of active markets, before the next market becomes active, the marginal increase in supply will be allocated in proportion to $1 / \demrate_j$ (in order to keep the prices the same).  This fully describes the behavior of the platform.

Without loss of generality, assume the markets are ordered such that $\demlim_1 \geq \ldots \geq \demlim_m$.  From the firm's point of view, the platform is equivalent to a single market with a piecewise linear demand curve: when the price is between $\demlim_1$ and $\demlim_2$, the rate at which price drops when $\supply$ increases is $\demrate_1$; for $\price \in [\demlim_2, \demlim_3]$, the rate is $1/(1 / \demrate_1 + 1 / \demrate_2)$. 
In general, when the first $k$ markets are active, prices drop i at the rate of $1/(\sum_{j = 1}^k 1/{\demrate_j})$.  We call this single demand curve the \emph{aggregate demand curve}.  

For a given production level, the area under the aggregate demand curve is equal to the welfare in the original markets.  The aggregate demand curve fully characterizes the set of markets under controlled access.  Note that whenever a new market joins, the rate at which price drops becomes slower, therefore the aggregate demand curve is always convex.

Conversely, one can show that any convex, decreasing, piecewise linear demand curve consisting of finitely many linear segments is equivalent to a set of linear demand markets under controlled access. \Halmos

\endproof

\proof{Proof of Theorem \ref{thm:ineff}.}
By \autoref{lem:agg-demand}, we can focus on constructing an aggregate demand curve.  Fix a constant $\zeta \in (0, \tfrac 1 2)$.  The aggregate demand curve we construct for $m$~markets (which is also the price function), $(m \geq 2)$, is
\begin{align*}
\price (\demand) = \max_{0 \leq k < m} (\zeta^k - \zeta^{2k} \demand).
\end{align*}
It is not hard to verify that this is the piecewise linear function that connects the following points:
\begin{align*}
    (0, 1), \left(\frac 1 {1 + \zeta}, \frac \zeta {1 + \zeta}\right), \left(\frac 1 {\zeta(1+\zeta)}, \frac{\zeta^2} {1 + \zeta}\right), \cdots \left(\frac 1 {\zeta^{m - 2} (1 + \zeta)}, \frac {\zeta^{m - 1}} {1 + \zeta}\right),\left(\frac 1 {\zeta^{m-1}}, 0\right)
\end{align*} 
Being the maximum of a family of decreasing linear functions, $\price(\demand)$ is obviously a convex decreasing function.	
We first calculate the efficient social welfare, which in the case of costless producers, is just the area under $\price(\demand)$.  The trapezoid whose vertices are $(1/ ({\zeta^{k-1} (1 + \zeta)}), 0)$,
$(1/({\zeta^{k-1} (1 + \zeta)}), {\zeta^k}/({1 + \zeta}))$,
$(1/({\zeta^{k} (1 +\zeta)}), 0)$,
$(1/({\zeta^{k} (1 + \zeta)}), {\zeta^{k+1}}/({1 + \zeta}))$ has area
\begin{align*}
\frac 1 2 \left( \frac {1}{\zeta^k (1 + \zeta)} - \frac {1}{\zeta^{k-1} (1+ \zeta)} \right) \left(\frac {\zeta^k}{1 + \zeta} + \frac{\zeta^{k+1}}{1 +\zeta} \right) = \frac {1 - \zeta}{1+\zeta}.
\end{align*}
There are $m - 2$ such trapezoids under $\price(\demand)$, and therefore the socially efficient welfare is $\Omega(m)$.  On the other hand, the linear components of $\price(\demand)$ are designed so that producing on any of the linear segment gives a maximal profit of $1/4$ 
(for the $k$-th segment, the profit maximizing production level is ${1}/({2\zeta^{k-1}})$).  The firm is indifferent to best responding to any of the linear segments, and all the production levels $1/({2\zeta^{k}})$ for $k = 0, \ldots, m - 1$ are equilibrium points.  If we push the starting point of $\price(\demand)$ from $(0, 1)$ to $(0, 1 +\eps)$ for some $\eps > 0$, then producing ${1+\eps}/2$ (by responding only to the first market) will be the unique equilibrium, resulting in a social welfare of only $3(1+\eps)^2 / 8$.  Therefore the price of anarchy is $\Omega(m)$. \Halmos
\endproof


\subsection{Proof of Theorem \ref{thm:generalcap}}

\proof{Proof of Theorem \ref{thm:generalcap}:}



\sam{Consider the following three cases.}

\sam{\noindent \textbf{Case 1: \ } $\lambda \in [2/3, 1]$. }


\rone{In this case, the main point is that when the maximal willingness to pay is equivalent, the platform would rather allocate to the market with higher price elasticity. Therefore, a similar example to the consumer surplus case would work. In particular, consider a one firm, two market setup with the following producer cost function and inverse demand function: 
$$
C(x) = (\alpha - \beta)x, ~ p_1(d_1) = \alpha - \epsilon d_1,~ p_2(d_2) = \alpha - \beta d_2, 
$$
where $\alpha > \beta > \epsilon > 0$. Under this scenario, the platform allocates first to the market with the larger price elasticity, which is market $2$. To be precise, the allocation function is as follows: 
\begin{align*}
    A_\lambda(x) = 
    \begin{cases}
    [0,x] & \text{if } 0\leq x\leq \frac{\alpha}{\beta}\\
    [x- \frac{\alpha}{\beta},  \frac{\alpha}{\beta}] & \text{if }  \frac{\alpha}{\beta}\leq x \leq \frac{\alpha}{\beta} +  \frac{\alpha}{\epsilon}\\ 
    [\frac{\alpha}{\epsilon}, \frac{\alpha}{\beta}]& \text{otherwise.} 
    \end{cases}
\end{align*}
Correspondingly, the price function can be computed to be: 
\begin{align*}
    p(A_\lambda(x)) = 
    \begin{cases}
    \alpha - \beta x & \text{if } 0\leq x\leq \frac{\alpha}{\beta}\\
    \frac{\alpha (x - \frac{\alpha}{\beta}) - \epsilon (x - \frac{\alpha}{\beta})^2}{x} & \text{if }  \frac{\alpha}{\beta}\leq x \leq \frac{\alpha}{\beta} +  \frac{\alpha}{\epsilon}\\ 
    0 & \text{otherwise.} 
    \end{cases}
\end{align*}
The efficient production profile is $d_1 = \beta/\epsilon$ and $d_2 = 1$, with social welfare $\alpha(\frac{\beta}{\epsilon}+1)$. On the other hand, taking derivatives over the piecewise profit function and checking for stationarity conditions yield the following two possible stationary values on production, i.e., $x = 1/2$ or $x = \alpha/\beta + \beta /(2\epsilon)$, and one can check that the profit obtained from only participating in the first market is no less than the profit in participating in both if: 
$$
\epsilon \geq \frac{\beta^3}{\beta^2 + 4\alpha(\alpha-\beta)}. 
$$
Setting $\epsilon$ to its lowest limit yields social welfare at equilibrium $s$ to be 
$$\SW(s,\lambda) = 3\beta/8,$$ 
which is independent of $\alpha$, while the efficient social welfare (after substituting the value of $\epsilon$) can be written 
$$
\SW^* = \alpha \left(2+ \frac{4\alpha(\alpha - \beta)}{\beta^2}\right) > 2\alpha, 
$$
and therefore the price of anarchy $\rho$ is
$$
\rho(\lambda,M) > \frac{16\alpha}{3\beta}, 
$$
which can be unboundedly large as $\alpha$ grows. 
}

\sam{\noindent \textbf{Case 2: \ } $\lambda \in [1/2, 2/3)$. }

In this case we can design a series of markets for a costless firm with a resulting price of anarchy that grows in the number of markets accessed, similar to the social welfare maximizing platform ($\lambda=0.5$ case). \rone{This is because as $\lambda$ goes from $1/2$ to $2/3$, the platform moves to the next market at a slower rate. Essentially, this means that $p(A_{1/2}(x)) \geq p(A_{\lambda}(x))$, for all $\lambda \in (1/2,2/3)$. As such, the same example for the case of $\lambda=1/2$ would yield the same price of anarchy results. We include a formal proof for $\lambda \in [1/2, 2/3)$ in the next case. }

\sam{\noindent \textbf{Case 3: \ } $\lambda \in (0, 1/2)$. }

\rone{We use a slightly modified example to show that} our price of anarchy lower bound decreases gracefully as the market maker objective approaches revenue maximizing, up till a point where it is worst to set up an open access platform, explaining the $\max$ operator. 

We define the set of production quantities which satisfy stationarity conditions on the profit function as the set of possible equilibrium points. We first show that regardless of the platform objective and the corresponding prices, the set of possible equilibrium points are uniquely defined.
\rone{\begin{lemma}\label{lem:allocationm}
	Regardless of the parameter $ \lambda $ for one costless firm, the quantity that satisfies stationarity (assuming the quantity is obtained before the platform allocates to the $(m+1)$-th market) when the platform allocates non-trivially to $m$ many markets is unique, i.e., $q_m^* = \sum_{j\in M} {\alpha_j}/({2\beta_j})$.
\end{lemma}}

\proof{Proof of Lemma \ref{lem:allocationm}:}
Since market $m$ is included, then the markets are allocated at least the following amount of quantity before the last market is accessed: 
\rone{$$
d_j = \frac{\alpha_j - \alpha_m}{ \left(\frac{2-3\lambda}{1-\lambda}\right)\beta_j}.
$$}
Further, we know that any additional quantity $Q$ introduced into the platform will be assigned $e_j$ to the respective markets respecting the following condition: 
$$
\beta_j e_j = \beta_m e_m \Rightarrow e_j = \frac{\frac{1}{\beta_j}}{\sum_k \frac{1}{\beta_k}}Q
$$ 
Optimizing over the additional quantity $e_j$, the firm's profit can be written: 
$$
\sum_{j=1}^m \left(\frac{\alpha_j - \alpha_m}{\sigma\beta_j}+e_j\right)\left( \alpha_j - \beta_j \left(\frac{\alpha_j - \alpha_m}{\sigma\beta_j}+e_j\right)\right)
$$ 
but with $e_j$ defined above and letting $B=\sum_k \frac{1}{\beta_k}$, this can be rewritten as: 
$$
\sum_{j=1}^m \left(\frac{\alpha_j - \alpha_m}{\sigma\beta_j}+\frac{Q}{B\beta_j}\right)\left( \alpha_j -\frac{\alpha_j - \alpha_m}{\sigma}-\frac{Q}{B}\right)
$$
Differentiating with respect to $Q$, we have: 
$$
\sum_{j=1}^m  \left[\frac{1}{B\beta_j}\left(\alpha_j - \frac{\alpha_j - \alpha_m}{\sigma} - \frac{Q}{B}\right) + \left(\frac{\alpha_j-\alpha_m}{\sigma\beta_j} + \frac{Q}{B\beta_j}\right)\left(-\frac{1}{B}\right)\right] = 0
$$
which gives us that: 
\rone{$$
Q = \sum_{j=1}^m \left(\frac{\alpha_j}{2\beta_j} - \frac{\alpha_j - \alpha_m}{\left(\frac{2-3\lambda}{1-\lambda}\right)\beta_j}\right),
$$}
implying that the aggregate production is indeed $\sum_{j=1}^m {\alpha_j}/({2\beta_j})$.
\Halmos
\endproof

We propose a set of parameters for each market as in the social welfare maximizing case and prove that the inefficiency in these markets are reflected as in Theorem \ref{thm:generalcap}. \rone{The inverse demand function for each market is as follows:}
\rone{$$
p_1(d_1) = 1 -d_1, \ \text{and} \ p_j(d_j) = \frac{a^{j-1}\theta^{j-1}}{1+\theta}- \frac{\theta^{2j-2}}{1-\theta^2}d_j, ~\forall j\geq 2,
$$
then as long as we can show that the (single costless) firm wants to participate only up till the first market, then the price of anarchy grows as a function of the sum of a geometric progression with ratio $a^2$, and thereby can be written: 
$$
\rho = \frac{4}{3}\left(1 + \frac{1-\theta}{1+\theta}\left(a^2 + a^4 + \dots \right)\right) = \frac{4}{3}\left(1+ \frac{(1-\theta)a^2}{(1+\theta)(1-a^2)}\right) 
$$
The optimal revenue considering only the first market is ${\alpha_1^2}/({4\beta_1})={1}/{4}$. On the other hand, if we assume that $m$ many markets are active, the amount allocated to market $j$ at the possible equilibrium point, denoted $d_{j,m}$, can be expressed as: 
$$
d_{j,m} =\left(\left(\frac{1}{2}-\frac{{1-\lambda}}{2-3\lambda}\right) \sum_{i=1}^m \frac{\alpha_i}{\beta_i}\right) \frac{\frac{1}{\beta_j}}{\sum_{i=1}^m \frac{1}{\beta_i}} + \frac{\alpha_j({1-\lambda})}{\beta_j\left(2-3\lambda\right)}
$$
and the prices at the markets then can be computed via $\alpha_j - \beta_j d_{j,m}$: 
$$
p_j(d_{j,m})=\left(1-\frac{1-\lambda}{{2-3\lambda}}\right)\alpha_j-\frac{1}{\sum_{i=1}^m \frac{1}{\beta_i}}\left(\left(\frac{1}{2}-\frac{{1-\lambda}}{2-3\lambda}\right) \sum_{i=1}^m \frac{\alpha_i}{\beta_i}\right)  
$$
and one can compute that revenue from all markets can be computed: 
$$
\left(\frac{1}{2} - \frac{1-\lambda}{2-3\lambda}\right)^2\frac{\left(\sum_{i=1}^m \frac{\alpha_i}{\beta_i}\right)^2}{\sum_{i=1}^m \frac{1}{\beta_i}} + \frac{1-\lambda}{2-3\lambda}\left(1-\frac{1-\lambda}{2-3\lambda}\right)\sum_{i=1}^m\frac{\alpha_i^2}{\beta_i}
$$
and we wish to show that $\forall \lambda$, we can find $a, ~\theta$ such that the above aggregate revenue is less than that when only the first market is active. (Recall that when $\lambda=1/2$, the second term is gone and the first term is the revenue from the social welfare case, and taking $a=1$, we retrieve the same right-hand side as the social welfare case too, which equates to the revenue at the first market, i.e., $1/4$. ) Further, recall from the $\lambda = 1/2$ case, that when we set $a=1$, we retrieve
$$
\frac{\left(\sum_{i=1}^m \frac{\alpha_i}{\beta_i}\right)^2}{\sum_{i=1}^m \frac{1}{\beta_i}} = 1
$$
Another important and interesting point to note is that the two coefficients in the revenue term sum up to $1/4$, i.e.,
\begin{equation*}
\left(\frac{1}{2} - \frac{1-\lambda}{2-3\lambda}\right)^2 +  \frac{1-\lambda}{2-3\lambda}\left(1-\frac{1-\lambda}{2-3\lambda}\right) = \frac{1}{4}.
\end{equation*}
We use this identity together with the fact that $(1/2 - (1-\lambda)/(2-3\lambda))^2$ increases in $\lambda$ (implying that the coefficient for the second term is negative) to prove the following lemma which confirms that the price of anarchy for the $\lambda \in [1/2,2/3)$ case is indeed at least the same as the $\lambda=1/2$ case: 
\begin{lemma}\label{largepoa_lemma}
For $\lambda\in [1/2,2/3)$, take $a=1$, and $\theta<0.5$, then 
\begin{equation*}
\left(\frac{1}{2} - \frac{1-\lambda}{2-3\lambda}\right)^2\frac{\left(\sum_{i=1}^m \frac{\alpha_i}{\beta_i}\right)^2}{\sum_{i=1}^m \frac{1}{\beta_i}} + \frac{1-\lambda}{2-3\lambda}\left(1-\frac{1-\lambda}{2-3\lambda}\right)\sum_{i=1}^m\frac{\alpha_i^2}{\beta_i} \leq \frac{1}{4} 
\end{equation*}
\end{lemma} 
\proof{Proof of Lemma \ref{largepoa_lemma}:}
From above, it remains to show that $\sum_{i=1}^m{\alpha_i^2}/{\beta_i} \geq 1$, but the first term in the sum is already $1$ while the rest are nonnegative so we are done. This implies that the revenue for all remaining possible stationary points is less than that when only the first market is active. \Halmos
\endproof
We now prove the case \rone{$\lambda<1/2$}. Note that $\alpha_i^2/\beta_i = 1 + ((1-\theta)/(1+\theta))\cdot[a^2 + a^4+ \dots]>1$, and therefore, as $\lambda$ approaches $0$, we also need to bring $a$ as close to $0$ as possible. For the convenience of the proof, we will use its limit as $j$ goes to infinity, i.e., $1 + ((1-\theta)/(1+\theta))\cdot[a^2/(1-a^2)]$. 
One can show that the first term can be computed to be for each $m$: 
$$
\left(\theta^{m-1} + \frac{a(1-\theta)(\theta^{m-1}-a^{m-1})}{\theta - a}\right)^2,
$$
and therefore it suffices to show the values for $a$ for each corresponding value of $\lambda$ for which  
$$
\left(\frac{1}{2} - \frac{1-\lambda}{2-3\lambda}\right)^2\left(\theta^{m-1} + \frac{a(1-\theta)(\theta^{m-1}-a^{m-1})}{\theta - a}\right)^2 + \frac{1-\lambda}{2-3\lambda}\left(1-\frac{1-\lambda}{2-3\lambda}\right)\left(1 + \frac{1-\theta}{1+\theta}\cdot \frac{a^2}{(1-a^2)}\right) \leq \frac{1}{4}.
$$
One can show that the first term (when setting $\theta$ sufficiently small, i.e., consider the case when $\theta \rightarrow 0$) can be bounded above by $a^2$, while the second term can be bounded by $1/(1-a^2)$. It therefore suffices to show that 
$$
\left(\frac{1}{2} - \frac{1-\lambda}{2-3\lambda}\right)^2a^2 + \frac{1-\lambda}{2-3\lambda}\left(1-\frac{1-\lambda}{2-3\lambda}\right)\left( \frac{1}{(1-a^2)}\right) \leq \frac{1}{4}.
$$
Define $\zeta = \left(\frac{1}{2} - \frac{1-\lambda}{2-3\lambda}\right)^2$, and $\eta = a^2$, then the above is but a quadratic inequality
$$
\zeta\cdot \eta + (\frac{1}{4} - \zeta)\left( \frac{1}{1-\eta}\right) \leq \frac{1}{4},
$$
which one can easily solve to find that the above holds when: 
$$
\eta \leq \frac{\left(\frac{1}{4} + \zeta\right) - \sqrt{(\zeta + 1/4)^2 - (2\zeta)^2}}{2\zeta},
$$
and substituting the original values we obtain
$$
a^2 \leq \frac{\left(\frac{1}{4} + \left(\frac{1}{2} - \frac{1-\lambda}{2-3\lambda}\right)^2\right) - \sqrt{\left(\left(\frac{1}{2} - \frac{1-\lambda}{2-3\lambda}\right)^2 + \frac{1}{4}\right)^2 - \left(2\left(\frac{1}{2} - \frac{1-\lambda}{2-3\lambda}\right)^2\right)^2}}{2\left(\frac{1}{2} - \frac{1-\lambda}{2-3\lambda}\right)^2},
$$
and after simplifying by setting $\zeta = \lambda^2/(4-6\lambda)^2$, we get 
$$
a^2 \leq \frac{(2-3\lambda)^2 + \lambda^2 - 4\sqrt{(1-\lambda)(1-2\lambda)(3\lambda^2-3\lambda+1)}}{2\lambda^2}.
$$
With that, one obtains the following lemma: 
\begin{lemma}
For $\lambda<1/2$, $a \leq \frac{1}{\sqrt{2}\lambda}\sqrt{(2-3\lambda)^2 + \lambda^2 - 4\sqrt{(1-\lambda)(1-2\lambda)(3\lambda^2-3\lambda+1)}}$ and taking $\theta$ small enough, then 
\begin{equation*}
\left(\frac{1}{2} - \frac{1-\lambda}{2-3\lambda}\right)^2\frac{\left(\sum_{i=1}^m \frac{\alpha_i}{\beta_i}\right)^2}{\sum_{i=1}^m \frac{1}{\beta_i}} +  \frac{1-\lambda}{2-3\lambda}\left(1-\frac{1-\lambda}{2-3\lambda}\right)\sum_{i=1}^m\frac{\alpha_i^2}{\beta_i} < \frac{1}{4}.
\end{equation*}
\end{lemma} 
}
The proof of Theorem \ref{thm:generalcap} follows from the above lemma, and the PoA bounds follow.

\section{Proofs from Section \ref{sec:dap}}

\subsection{Proof of Theorem \ref{thm:greedy-alg}} 
Without loss of generality, we assume that $c_1 \leq \cdots \leq c_n$. 
We only provide a proof for the case in which the number of market $m = 1$,
as it is straightforward to generalize our proof to the case in which $m > 1$ under the assumption on the linearity of firms' cost functions.


We denote by $F_1 (\Ecal)$ the set of firms that have access to market $1$.
It is defined according to
$$F_1 (\Ecal) = \{ i \in F \, | \, (i,1) \in \Ecal \}. $$
We first introduce the concept of a contiguous set of firms, which plays a central role in the remainder of the proof. We have the following definition.
\begin{definition}[Contiguous Set]
	The set $F_1 (\Ecal)$ is \emph{contiguous} if
	$$F_1 (\Ecal) = \{1, 2, \dots, |F_1 (\Ecal)| \}, \quad \text{or} \quad F_1 (\Ecal) = \emptyset.$$ 
\end{definition}
\noindent Here, $|F_1 (\Ecal)|$ denotes the cardinality of the set $F_1 (\Ecal)$.
Qualitatively, the set $F_1 (\Ecal) \subseteq F$ is contiguous, if it consists of consecutive elements of the set $F$.
Clearly, for the edge set $\Ecal^*$ generated by the greedy algorithm, the set $F_1 (\Ecal^*)$ is contiguous. 

The rest of the proof consists of two parts. In Part 1, we show that if the set $F_1 (\Ecal)$ is contiguous, then the social welfare at the Nash equilibrium associated with edge set $\Ecal$ is guaranteed to be no larger than that of the edge set $\Ecal^*$. In Part 2, we consider the case in which the set $F_1 (\Ecal)$ is not contiguous. We show that there exists an edge set $\widetilde{\Ecal}$ that yields a contiguous set $F_1 (\widetilde{\Ecal})$, and has social welfare at Nash equilibrium that is no smaller than that of the edge set $\Ecal$.

\vspace{.05in}

\noindent \emph{Part 1:} 
In this part, we assume that the set $F_1 (\Ecal)$ is contiguous, and show that $\SW (q^{\rm NE} (\Ecal) , C) \leq \SW (q^{\rm NE} (\Ecal^*) , C) $.
We first define a sequence of edge sets according to
\begin{align}
\Ecal_k = \bigcup_{i=1}^k \{(i,1)\}, \quad k = 0, \dots, n. \label{eq:Ek}
\end{align}
In particular, we have that $\Ecal_0 = \emptyset$. Let $k^* = |\Ecal^*|$. 
To show that $\SW (q^{\rm NE} (\Ecal) , C) \leq \SW (q^{\rm NE} (\Ecal^*) , C) $ if $F_1 (\Ecal)$ is contiguous, it suffices to show that the sequence $\SW \left( q^{\rm NE} (\Ecal_k) , \,  C  \right)$ is strictly increasing in $k$ over $ k = 0, \dots, k^*$, and monotonically non-increasing in $k$ over $k = k^* , \dots, n$. 

We assume without loss of generality that $c_i \leq \alpha_1$ for all $i$. If this is not the case, one can work with an alternative cost function profile  $\widetilde{C} = ( \widetilde{C}_1, \dots, \widetilde{C}_n )$ that is defined according to 
$$\widetilde{C}_i (s_i) = \left( \min \{ c_i, \alpha_1\} \cdot s_i \right)^+$$
for $i = 1, \dots, n$. Clearly, $\min \{c_i, \alpha_1\} \leq \alpha_1$ for all $i$. Additionally, it is straightforward to show that
$$\SW (q^{\rm NE} (\Ecal_k), C) = \SW (q^{\rm NE} (\Ecal_k), \widetilde{C})$$
for all $k = 0, \dots, n$.


The proof of this claim on monotonicity relies on the following lemma. Its proof is deferred to Appendix \ref{pf:lem:pf_greedy1}.

\begin{lemma} \label{lem:pf_greedy1}
	Assume that each firm $i$'s cost function is of the form $C_i (s_i) = (c_i s_i)^+$, where $c_1 \leq \cdots \leq c_n$.
	Let the number of markets $m = 1$, and the edge set $\Ecal_k$ be defined according to Eq. \eqref{eq:Ek}. For each $k = 1, \dots, n$, we have that $\SW \left( q^{\rm NE} (\Ecal_k) , \,  C  \right) > \SW \left( q^{\rm NE} (\Ecal_{k-1}) , \, C \right)$ if and only if
	\begin{align}
	\alpha_1 - c_k > \frac{1}{k} \left( 1 + \frac{1}{k - \frac{1}{2(k+1)}} \right) \left( \sum_{i=1}^{k-1} (\alpha_1 - c_i) \right). \label{eq:cond_inc_SW}
	\end{align}
\end{lemma}

It follows from the description of the greedy algorithm that $\SW \left( q^{\rm NE} (\Ecal_k) , \,  C  \right)$ is strictly increasing in $k$ for $0 \leq k \leq k^*$. Additionally, we have that 
$$\SW \left( q^{\rm NE} (\Ecal_{k^*+1}) , \,  C  \right) \leq \SW \left( q^{\rm NE} (\Ecal_{k^*}) , \, C \right).$$
It follows from Lemma \ref{lem:pf_greedy1} that for $k = k^* + 1$, the following inequality is satisfied:
\begin{align}
\alpha_1 - c_k \leq \frac{1}{k} \left( 1 + \frac{1}{k - \frac{1}{2(k+1)}} \right) \left( \sum_{i=1}^{k-1} (\alpha_1 - c_i) \right). \label{eq:cond_dec_SW}
\end{align}
To complete Part 1 of the proof, we have the following lemma. Its proof can be found in Appendix \ref{pf:lem:pf_greedy2}.

\begin{lemma} \label{lem:pf_greedy2}
	Let $k^* \in \{0, \dots, n \}$, and assume that $c_1 \leq \cdots \leq c_n \leq \alpha_1$.  If inequality \eqref{eq:cond_dec_SW} is satisfied for $k = k^* +1$,
	then it is satisfied for $k = k^* + 1, \dots, n$.
\end{lemma}
\noindent A combination of Lemma \ref{lem:pf_greedy1} and \ref{lem:pf_greedy2} reveals that
$$\SW \left( q^{\rm NE} (\Ecal_{k}) , \,  C  \right) \leq \SW \left( q^{\rm NE} (\Ecal_{k-1}) , \, C \right)$$
for $k = k^* + 1, \dots, n$. This completes Part 1 of the proof.

\vspace{.05in}

\noindent \emph{Part 2:} In this part, we assume that the set $F_1 (\Ecal)$ is not contiguous. We show that there exists an edge set $\widetilde{\Ecal} \subseteq F \times \{1\}$, such that the set $F_1 (\widetilde{\Ecal})$ is contiguous, and $\SW \left( q^{\rm NE} (\Ecal), \, C \right) \leq \SW \left( q^{\rm NE} (\widetilde{\Ecal}) , \, C \right)$.

Our proof of the above claim is constructive. Given an edge set $\Ecal \subseteq F \times \{1\}$, we construct a sequence of $(n^2+1)$ edge sets according to the following procedure
\begin{enumerate}
	\item Set $k = 0$, and $\Ecal_k = \Ecal$.
	\item If the set $F_1 (\Ecal_k)$ is contiguous, then set $\Ecal_{k+1} = \Ecal_k$, and go to Step 5. If not, go to Step 3.
	\item Define the edge set $\widetilde{\Ecal}_k$ according to
	\begin{align}
	\widetilde{\Ecal}_k = \Ecal_k \setminus \left\{ \big( \max F_1 (\Ecal_k), 1 \big) \right\}. \label{eq:tilde_E_t}
	\end{align}
	\item If $\SW \left( q^{\rm NE} (\Ecal_k), \, C \right) \leq \SW \left( q^{\rm NE} (\widetilde{\Ecal}_k) , \, C \right)$, then set $\Ecal_{k+1} = \widetilde{\Ecal}_k$. If not, set
	\begin{align}
	\Ecal_{k+1} = \widetilde{\Ecal}_k \cup \left\{ \left( \min  \left( F \setminus F_1 (\Ecal_k) \right), 1 \right) \right\}. \label{eq:E_t+1}
	\end{align}
	\item If $k < n^2$, update $k = k+1$, and go to Step 2. If $k \geq n^2$, terminate the procedure.
\end{enumerate}
We will show that both of the following claims are true
\begin{enumerate}[(i)]
	\item The set $F_1 (\Ecal_{n^2})$ is contiguous.
	\item For $k = 0, \dots, n^2 -1 $, we have
	\begin{align}
	\SW \left( q^{\rm NE} (\Ecal_k), \, C \right) \leq \SW \left( q^{\rm NE} (\Ecal_{k+1}) , \, C \right). \label{eq:inc_SW}
	\end{align}
\end{enumerate}
We  note that the second claim implies that $\SW (q^{\rm NE} (\Ecal), C) \leq \SW (q^{\rm NE}(\Ecal_{n^2}), C)$.
In what follows, we show that Claim (i) and (ii) are true in Part 2.1 and 2.2 of the proof, respectively.

\vspace{.05in}

\noindent \emph{Part 2.1: Proof of Claim (i).}
We show that Claim (i) is true according to a ``potential function" argument. Namely, we define a potential function on the edge set $\Ecal_k \subseteq F \times \{1\}$ as follows
$$\Phi (\Ecal_k) = |F_1 (\Ecal_k)| \big( \max F_1 (\Ecal_k) - |F_1 (\Ecal_k)| \big)$$
for $k = 0, \dots, n^2$.  For all $\Ecal_k \subseteq F \times \{1\}$, we have that $\Phi (\Ecal_k) \geq 0$. It is straightforward to show $\Phi (\Ecal_k) = 0$ if and only if the set $F_1 (\Ecal_k)$ is contiguous. 
It follows that if $\Phi (\Ecal_k) > 0$, then $\Ecal_{k+1}$ is specified according to either $\Ecal_{k+1} = \widetilde{\Ecal}_k$ or Eq. \eqref{eq:E_t+1}.
If $\Ecal_{k+1} = \widetilde{\Ecal}_k$, we have
\begin{align}
&\Phi (\Ecal_{k+1} ) =  |F_1 (\widetilde{\Ecal}_k)| \big( \max F_1 (\widetilde{\Ecal}_k) - |F_1 (\widetilde{\Ecal}_k)| \big) \\
& \qquad =|F_1 (\widetilde{\Ecal}_k)| \big( \max F_1 (\widetilde{\Ecal}_k)  - \left( |F_1 (\Ecal_k)| -1 \right) \big) \\
& \qquad \leq |F_1 (\widetilde{\Ecal}_k)| \big( \max F_1 (\Ecal_k) -1  - \left( |F_1 (\Ecal_k)| -1 \right) \big) \label{eq:step3_1}\\
& \qquad = \left( |F_1 (\Ecal_k)| - 1 \right) \big( \max F_1 (\Ecal_k)   -  |F_1 (\Ecal_k)| \big) \label{eq:step3_2}\\
& \qquad \leq \Phi (\Ecal_k) - 1, \label{eq:step3_3}
\end{align}
where inequality \eqref{eq:step3_1} follows from the inequality  $\max F_1 (\widetilde{\Ecal}_k ) \leq \max F_1 (\Ecal_k) - 1$, Eq. \eqref{eq:step3_2} follows from the fact that $|F_1 (\widetilde{\Ecal}_k)| =  |F_1 (\Ecal_k)| - 1$, and inequality \eqref{eq:step3_3} follows from the fact that $F_1 (\Ecal_k)$ is not contiguous.

On the other hand, if $\Ecal_{k+1}$ is specified according to Eq. \eqref{eq:E_t+1}, then we have
\begin{align}
\Phi (\Ecal_{k+1}) = &|F_1 (\Ecal_k)| \big( \max F_1 (\Ecal_{k+1}) - |F_1 (\Ecal_k)| \big) \label{eq:step4_0}\\
\leq & |F_1 (\Ecal_k)| \big( \max F_1 (\Ecal_k) - 1 - |F_1 (\Ecal_k)| \big) \label{eq:step4} \\
= &\Phi (\Ecal_k) - |F_1 (\Ecal_k)| \leq \Phi (\Ecal_k) - 1,
\end{align}
where Eq. \eqref{eq:step4_0} follows from the fact that $|F_1 (\Ecal_k) | = |F_1 (\Ecal_{k+1} )|$, and inequality \eqref{eq:step4} follows from the fact that
$\max F_1 (\Ecal_{k+1} ) \leq \max F_1 (\Ecal_k) - 1$. In both cases, we have that if $\Phi (\Ecal_k) > 0$, then
$$\Phi (\Ecal_k) - \Phi (\Ecal_{k+1} ) \geq 1.$$
It is straightforward to show that $\Phi (\Ecal_0) < n^2$. It immediately follows that $\Phi (\Ecal_{n^2} ) = 0$. This finishes the proof of Claim (i).

\vspace{.05in}

\noindent \emph{Part 2.2: Proof of Claim (ii).}
According to the procedure we use in generating the sequence of edge sets, inequality \eqref{eq:inc_SW} is trivially satisfied if $\Ecal_{k+1} = \Ecal_k$ or $\Ecal_{k+1} = \widetilde{\Ecal}_k$.
%
For the remainder of this part, we show that inequality \eqref{eq:inc_SW} is satisfied if $\Ecal_{k+1}$ is specified according to Eq. \eqref{eq:E_t+1}.
The key idea in this proof is to show that if the removal of the most expensive producer leads to a strict decrease in social welfare, then a unilateral decrease in the marginal cost of said producer will lead to an increase in social welfare.

We first introduce some notation pertinent to the remainder of the proof.
Define the indices $g_k$ and $h_k$ according to 
\begin{align*}
g_k = \max F_1 (\Ecal_k) \quad \text{and} \quad h_k = \min \left( F \setminus F_1 (\Ecal_k) \right).
\end{align*}
Since the set $F_1 (\Ecal_k)$ is not contiguous, we must have that $h_k < g_k$.
We define a new cost function profile $C^{\theta} = (C_1^{\theta}, \dots, C_n^{\theta})\in \Lcal^n (c_{\min}, c_{\max})$ according to 
\begin{align}
C_i^{\theta} (s_i) = \begin{cases}
(c_i s_i)^+ & \text{if } i \neq g_k \\
(\theta s_i)^+ & \text{if } i = g_k, 
\end{cases} \label{eq:C_theta}
\end{align}
where $\theta$ is a scalar parameter.

%

The unique Nash equillibrium of the networked Cournot game depends on firms' cost function profiles. 
With a slight abuse of notation, we denote by $q^{\rm NE} (\Ecal , C)$ the unique Nash equilibrium of the networked Cournot game associated with an edge set $\Ecal \subseteq F \times \{1\}$, and a cost function profile $C$.

Our proof relies on the following technical lemma on the monotonicity of $\SW (q^{\rm NE} (\Ecal_k, C^{\theta}), C^{\theta})$ in the scalar $\theta$. Its proof  is deferred to Appendix \ref{pf:lem:pf_greedy3}.

\begin{lemma}\label{lem:pf_greedy3}
	Let the edge set $\widetilde{\Ecal}_k$ and the cost function profile $C^{\theta}$ be specified according to Eq. \eqref{eq:tilde_E_t} and \eqref{eq:C_theta}, respectively. If $\SW \left( q^{\rm NE} (\Ecal_k, C), \, C \right) > \SW \left( q^{\rm NE} (\widetilde{\Ecal}_k, C) , \, C \right)$, then $\SW (q^{\rm NE} (\Ecal_k, C^{\theta}), C^{\theta})$ is monotonically decreasing in $\theta$ for $c_{\min} \leq \theta \leq c_{g_k}$.
\end{lemma}


Note that, by specifying $\Ecal_{k+1}$ according to Eq. \eqref{eq:E_t+1}, we essentially replace the most expensive producer with another producer with a cheaper cost.
In particular, if $\Ecal_{k+1}$ is specified according to Eq. \eqref{eq:E_t+1}, then $\SW (q^{\rm NE} (\Ecal_k, C), C)$ and $\SW (q^{\rm NE} (\Ecal_{k+1}, C), C)$ are related according to
\begin{align}
\SW (q^{\rm NE} (\Ecal_k, C), C) &= \SW (q^{\rm NE} (\Ecal_k, C^{c_{g_k}}), C^{c_{g_k}}), \label{eq:SW_k}\\
\SW (q^{\rm NE} (\Ecal_{k+1}, C), C) &= \SW (q^{\rm NE} (\Ecal_k, C^{c_{h_k}}), C^{c_{h_k}}) . \label{eq:SW_k+1}
\end{align}
Recall that $h_k < g_k$. It follows that $c_{\min} \leq c_{h_k} \leq c_{g_k}$.
An application of Lemma \ref{lem:pf_greedy3} shows that
\begin{align*}
\SW (q^{\rm NE} (\Ecal_k, C^{c_{g_k}}), C^{c_{g_k}}) 
\leq \SW (q^{\rm NE} (\Ecal_k, C^{c_{h_k}}), C^{c_{h_k}}) .
\end{align*}
The above inequality, in combination with Eq. \eqref{eq:SW_k} and \eqref{eq:SW_k+1}, shows that when $\Ecal_{k+1}$ is specified according to Eq. \eqref{eq:E_t+1}, inequality \eqref{eq:inc_SW} is  satisfied. This completes the proof.

\subsection{Proof of Theorem \ref{thm:discriminatory_PoA}}
\noindent Without loss of generality, we assume that $c_1 \leq \cdots \leq c_n$.
It follows from the description of the greedy algorithm that for each market $j \in M$, $(1, j) \in \Ecal^*$ if and only if $c_1 < \alpha_j$. It immediately follows that the efficient social welfare associated with the edge set $\Ecal^*$ satisfies
\begin{align*}
\SW^* (\Ecal^*, C) = \sum_{j=1}^m  \frac{\left( (\alpha_j - c_1 )^+ \right)^2}{2 \beta_j} = \SW^* (F\times M, C). 
\end{align*}

For the second part of the theorem, we only provide a proof for the case in which the number of markets $m = 1$. 
The generalization to the case in which $m > 1$ can be carried out following similar steps as in the proof of \cref{prop:PoA_cmin_cmax}. 

Given the restriction that $c_1 \leq \cdots \leq c_n$, we define the subset of ordered linear cost function profiles in $\Lcal^n (c_{\min}, c_{\max})$ according to
\begin{align*}
&\Ocal^n (c_{\min}, c_{\max}) 
= \bigg\{ C \in \Lcal^n (c_{\min}, c_{\max})  \, \bigg|  \, C_i(s_i ) = (c_i s_i)^+, \; i = 1, \dots, n , \; c_1 \leq \cdots \leq c_n \bigg\}.
\end{align*}
Thus, we have the following chain of inequalities
\begin{align}
&\rho (\Ecal^* , C) = \inf_{\Ecal \subseteq F \times M} \ \frac{\SW^* (F\times M, C ) }{\SW (q^{\rm NE} (\Ecal), C)} \label{eq:pf_disc_1} \\
\leq &\sup_{C \in \Lcal^n (c_{\min}, c_{\max})} \ \ \inf_{\Ecal \subseteq F \times M} \frac{\SW^* (F\times M, C ) }{\SW (q^{\rm NE} (\Ecal), C)} \\
= &\sup_{C \in \Ocal^n (c_{\min}, c_{\max}) } \ \  \inf_{\{(1,1)\} \subseteq \Ecal \subseteq F \times M} \frac{\SW^* (F\times M, C ) }{\SW (q^{\rm NE} (\Ecal), C)} \label{eq:pf_disc_3} \\
\leq & \inf_{\{(1,1)\} \subseteq \Ecal \subseteq F \times M} \  \ \sup_{C \in \Ocal^n (c_{\min}, c_{\max}) } \frac{\SW^* (F\times M, C ) }{\SW (q^{\rm NE} (\Ecal), C)} \label{eq:pf_disc_4}\\
\leq & \inf_{\{(1,1)\} \subseteq \Ecal \subseteq F \times M} \ \ \frac{1}{\frac{2 |\Ecal| + 4}{3 |\Ecal + 5|} + \delta (\gamma_1, |\Ecal|)} \label{eq:pf_disc_5} \\
= &\frac{1}{\underset{k \in \{1, \dots, n\}}{\max} \left\{ \frac{2k+4}{3k+5}  + \delta (\gamma_1, k) \right\} }.
\end{align}
Here, Eq. \eqref{eq:pf_disc_1} follows from the fact that $\SW^* (\Ecal^*, C) = \SW^* (F \times M, C)$, Eq. \eqref{eq:pf_disc_3} follows from our restriction that $c_1 \leq \cdots \leq c_n$, inequality \eqref{eq:pf_disc_4} follows from the min-max inequality, and inequality \eqref{eq:pf_disc_5} is a direct application of Proposition \ref{prop:PoA_cmin_cmax}.
This completes the proof.

\subsection{Proof of Lemma \ref{lem:pf_greedy1}} \label{pf:lem:pf_greedy1}
\noindent 
The proof proceeds in two parts. In Part 1, we provide a necessary and sufficient condition for the production quantity of firm $k$ to be strictly positive at Nash equilibrium, when the edge set is given by $\Ecal_k$. In Part 2, we leverage on the intermediary result in Part 1 to show that
$\SW \left( q^{\rm NE} (\Ecal_k) , \,  C  \right) > \SW \left( q^{\rm NE} (\Ecal_{k-1}) , \, C \right)$ if and only if inequality \eqref{eq:cond_inc_SW} is satisfied.

\vspace{.05in}

\noindent \emph{Part 1:} We show that for the game $(F, \Qcal (\Ecal_k), \pi)$, the production quantity of firm $k$ at Nash equilibrium $q^{\rm NE}_{k1} (\Ecal_k)$ is strictly positive, if and only if 
\begin{align}
\alpha_1 - c_k > \frac{1}{k}  \left( \sum_{i=1}^{k-1} (\alpha_1 - c_i) \right). \label{eq:k_active}
\end{align}

We first prove the ``if" part of the desired claim. First note that inequality \eqref{eq:k_active} implies that the following inequality is satisfied:
\begin{align}
(k+1) (\alpha_1 - c_k)  > \sum_{i = 1}^k (\alpha_1 - c_i ). \label{eq:k_active_2}
\end{align}
One can check that if inequality \eqref{eq:k_active} is satisfied, then the unique Nash equilibrium of the game $(F, \Qcal (\Ecal_k), \pi)$ is given by
\begin{align*}
q_{i1}^{\rm NE} (\Ecal_k) = \frac{(k+1) (\alpha_1 - c_i)  - \sum_{\ell = 1}^k (\alpha_1 - c_{\ell} ) }{(k+1) \beta_1},  \ i = 1, \dots, k, 
\end{align*}
and $q_{i1}^{\rm NE} (\Ecal_k) = 0$ for $i = k+1, \dots, n$. It follows from inequality \eqref{eq:k_active_2} that $q_{k1}^{\rm NE} (\Ecal_k) > 0$. 

Next, we prove the ``only if" part of the desired claim. 
First note that $c_i \leq c_k$ for $i = 1, \dots, k$. Recall that for the game $(F, \Qcal (\Ecal_k), \pi)$, firm $k$'s production quantity $q^{\rm NE}_{k1} (\Ecal_k)$ at Nash equilibrium is strictly positive. It follows that $q^{\rm NE}_{i1} (\Ecal_k) > 0$ for $i = 1 ,\dots, k$. The first order optimality condition for Nash equilibrium of the game $(F, \Qcal (\Ecal_k), \pi)$ implies that
\begin{align}
\alpha_1 - \beta_1 \left( \sum_{\ell=1}^k q_{\ell 1}^{\rm NE} (\Ecal_k) \right) - \beta_1 q_{i1}^{\rm NE} (\Ecal_k) - c_i = 0,
\end{align}
for $i = 1, \dots, k$. Consequently, we have that
$$q_{k1}^{\rm NE} (\Ecal_k) = \frac{(k+1) (\alpha_1 - c_k)  - \sum_{\ell = 1}^k (\alpha_1 - c_{\ell} ) }{(k+1) \beta_1} > 0.$$
This implies that inequality \eqref{eq:k_active} is satisfied.

\vspace{.05in}

\noindent \emph{Part 2:} We show  $\SW \left( q^{\rm NE} (\Ecal_k) , \,  C  \right) > \SW \left( q^{\rm NE} (\Ecal_{k-1}) , \, C \right)$ if and only if inequality \eqref{eq:cond_inc_SW} is satisfied.
First note when $k = 1$, it is straightforward to see that $\SW \left( q^{\rm NE} (\Ecal_1) , \,  C  \right) > 0$ if and only if $\alpha_1 - c_1 > 0$. Thus, for the remainder of the proof, we assume that $k \geq 2$.

We only provide a proof for the ``only if" part of the claim, as the ``if" part of the claim can be proved using similar arguments. First note that $\SW \left( q^{\rm NE} (\Ecal_k) , \,  C  \right) > \SW \left( q^{\rm NE} (\Ecal_{k-1}) , \, C \right)$ implies that $q_{k1}^{\rm NE} (\Ecal_k) > 0$. If this is not the case, then we have that $q^{\rm NE} (\Ecal_k) = q^{\rm NE} (\Ecal_{k-1})$, which clearly leads to a contradiction. Note that  $c_i \leq c_k$ for $i = 1, \dots, k$. It follows that 
$$q_{i1}^{\rm NE} (\Ecal_k) \geq q_{k1}^{\rm NE} (\Ecal_k) > 0, \quad \text{for} \quad i = 1, \dots, k.$$
One can show that the unique Nash equilibrium of the game $(F, \Qcal (\Ecal_k), \pi)$ is given by
\begin{align*}
q_{i1}^{\rm NE} (\Ecal_k) = \frac{(k+1) (\alpha_1 - c_i)  - \sum_{\ell = 1}^k (\alpha_1 - c_{\ell} ) }{(k+1) \beta_1},  \ i = 1, \dots, k, 
\end{align*}
and $q_{i1}^{\rm NE} (\Ecal_k) = 0$ for $i = k+1, \dots, n$. We denote the aggregate supply in market 1 at the unique Nash equilibrium of the game $(F, \Qcal (\Ecal_k), \pi)$ by $d_1^{\rm NE}(\Ecal_k)$. It is given by
\begin{align*}
d_1^{\rm NE} (\Ecal_k) = \sum_{i=1}^n q_{i1}^{\rm NE} (\Ecal_k) =  \frac{  \sum_{i = 1}^k (\alpha_1 - c_i ) }{(k+1) \beta_1}
\end{align*}
Additionally, the social welfare at the Nash equilibrium of the game $(F, \Qcal (\Ecal_k), \pi)$ satisfies
\begin{align*}
&\SW \left( q^{\rm NE} (\Ecal_k), C \right) \\
= &\alpha_1 d_1^{\rm NE} (\Ecal_k) - \frac{1}{2} \beta_1 {d_1^{\rm NE}}^2 (\Ecal_k) - \sum_{i=1}^k c_i q_{i1}^{\rm NE} (\Ecal_k) \\
= &\alpha_1 d_1^{\rm NE} (\Ecal_k) - \frac{1}{2} \beta_1 {d_1^{\rm NE}}^2 (\Ecal_k) - \sum_{i=1}^k c_i \left( \frac{\alpha_1 - c_i}{\beta_1} - d_1^{\rm NE} (\Ecal_k) \right) \\
= &(k+1) \alpha_1 d_1^{\rm NE} (\Ecal_k)  - \frac{1}{2} \beta_1 {d_1^{\rm NE}}^2 (\Ecal_k) - \sum_{i=1}^k \left(   \frac{c_i (\alpha_1 - c_i)}{\beta_1}  \right) + \left( -k \alpha_1 + \sum_{i=1}^k c_i \right) d_1^{\rm NE} (\Ecal_k) \\
= & \frac{\alpha_1  \sum_{i=1}^k (\alpha_1 - c_i)}{\beta_1}  - \frac{2k+3}{2} \beta_1 {d_1^{\rm NE}}^2 (\Ecal_k) - \sum_{i=1}^k  \frac{c_i (\alpha_1 - c_i)}{\beta_1}   \\
= &\frac{\sum_{i=1}^k (\alpha_1 - c_i)^2}{\beta_1} - \frac{2k+3}{2} \beta_1 {d_1^{\rm NE}}^2 (\Ecal_k) \\
= &\frac{\sum_{i=1}^k (\alpha_1 - c_i)^2}{\beta_1} - \frac{2k+3}{2} \frac{\left( \sum_{i=1}^k (\alpha_1 - c_i) \right)^2}{(k+1)^2 \beta_1}.
\end{align*}

Next, we provide a closed-form expression for $\SW \left( q^{\rm NE} (\Ecal_{k-1}), C \right)$. Recall that $q^{\rm NE}_{k1} (\Ecal_k) > 0$. As we showed in Part 1, this implies that inequality \eqref{eq:k_active} is satisfied. We thus have
$$\alpha_1 - c_{k-1} \geq \alpha_1 - c_k > \frac{1}{k}  \left( \sum_{i=1}^{k-1} (\alpha_1 - c_i) \right),$$
which further implies that
$$\alpha_1 - c_{k-1}  > \frac{1}{k-1}  \left( \sum_{i=1}^{k-2} (\alpha_1 - c_i) \right).$$
It follows from our result in Part 1 that for the game $(F, \Qcal (\Ecal_{k-1}), \pi)$, producer $k-1$'s production quantity at Nash equilibrium is strictly positive.
Using similar arguments as in our derivation on the closed-form expression for $\SW \left( q^{\rm NE} (\Ecal_{k}), C \right)$, we have the following closed-form expression for $\SW \left( q^{\rm NE} (\Ecal_{k-1}), C \right)$
%
\begin{align*}
&\SW \left( q^{\rm NE} (\Ecal_{k-1}), C \right) \\
& \qquad = \frac{\sum_{i=1}^{k-1} (\alpha_1 - c_i)^2}{\beta_1} - \frac{2k+1}{2} \frac{\left( \sum_{i=1}^{k-1} (\alpha_1 - c_i) \right)^2}{k^2 \beta_1}.
\end{align*}

Thus, the difference between $\SW \left( q^{\rm NE} (\Ecal_k) , \,  C  \right)$ and $\SW \left( q^{\rm NE} (\Ecal_{k-1}) , \, C \right)$ is given by
\begin{align*}
&\SW \left( q^{\rm NE} (\Ecal_k) , \,  C  \right) - \SW \left( q^{\rm NE} (\Ecal_{k-1}) , \, C \right)  \\
&\qquad =  \frac{(\alpha_1 - c_k)^2}{\beta_1} + \frac{2k+1}{2} \frac{\left( \sum_{i=1}^{k-1} (\alpha_1 - c_i) \right)^2}{k^2 \beta_1}   \\
&\qquad \phantom{=} - \frac{2k+3}{2}  \frac{\left( \sum_{i=1}^k (\alpha_1 - c_i) \right)^2}{(k+1)^2 \beta_1}.
\end{align*}
An algebraic calculation reveals that $\SW \left( q^{\rm NE} (\Ecal_k) , \,  C  \right) > \SW \left( q^{\rm NE} (\Ecal_{k-1}) , \, C \right)$ if and only if 
\begin{align*}
\frac{\alpha_1 - c_k}{\sum_{i=1}^{k-1} (\alpha_1 - c_i) } <  \frac{1}{k}, \quad \text{or } \  \frac{\alpha_1 - c_k}{\sum_{i=1}^{k-1} (\alpha_1 - c_i) } > \frac{k+2+ \frac{1}{2k}}{k^2 + k - \frac{1}{2}}.
\end{align*}
Recall that $\SW \left( q^{\rm NE} (\Ecal_k) , \,  C  \right) > \SW \left( q^{\rm NE} (\Ecal_{k-1}) , \, C \right)$ implies that inequality \eqref{eq:k_active} is satisfied. It follows from the above inequality that $\SW \left( q^{\rm NE} (\Ecal_k) , \,  C  \right) > \SW \left( q^{\rm NE} (\Ecal_{k-1}) , \, C \right)$ implies  inequality \eqref{eq:cond_inc_SW} is satisfied.


\subsection{Proof of Lemma \ref{lem:pf_greedy2}} \label{pf:lem:pf_greedy2}
\noindent We prove this lemma by induction in $k$. 

\vspace{.05in}

\noindent \emph{Base Step:} For $k = k^* + 1$, inequality \eqref{eq:cond_dec_SW} is satisfied by the assumption of this lemma.

\vspace{.05in}

\noindent \emph{Induction Step:}
Assume that inequality \eqref{eq:cond_dec_SW} is satisfied for $k \geq 1$. We show that it is satisfied for $k+1$ by showing that
\begin{align}
\alpha_1 - c_{k+1} \leq \frac{1}{k+1} \left( 1 + \frac{1}{k+1 - \frac{1}{2(k+2)}}\right)  \left( \sum_{i=1}^{k} (\alpha_1 - c_i)  \right).\label{eq:firm_k_plus_1}
\end{align}
Since $c_{k+1} \geq c_k$, inequality \eqref{eq:firm_k_plus_1} is satisfied if the following inequality holds
\begin{align}
(k + 1) ( \alpha_1 - c_k )  \leq  \left( 1 + \frac{1}{k+1 - \frac{1}{2(k+2)}}\right)  \left( \sum_{i=1}^{k} (\alpha_1 - c_i)  \right).  \label{eq:pf_lem_greedy2_ineq1}
\end{align}
An algebraic calculation reveals that inequality \eqref{eq:pf_lem_greedy2_ineq1} is satisfied if and only if
\begin{align*}
\frac{k^3 + 3k^2 + \frac{1}{2} k - 2}{k^2 + 3k + \frac{3}{2}} (\alpha_1 - c_k)  \leq \frac{ k^2 + 4 k + \frac{7}{2} }{k^2 + 3k + \frac{3}{2}} \left( \sum_{i=1}^{k-1} (\alpha_1 - c_i)  \right).
\end{align*}
Given that  $k \geq 1$, the above inequality is satisfied if and only if
\begin{align}
\alpha_1 - c_k \leq \frac{ k^2 + 4 k + \frac{7}{2} }{k^3 + 3k^2 + \frac{1}{2} k - 2} \left( \sum_{i=1}^{k-1} (\alpha_1 - c_i)  \right). \label{eq:pf_lem_greedy2_ineq2}
\end{align}
The induction hypothesis implies that inequality \eqref{eq:cond_dec_SW} is satisfied for $k$. Given that $c_i \leq \alpha_1$ for $i = 1, \dots, n$, we have that inequality \eqref{eq:pf_lem_greedy2_ineq2} is satisfied if
\begin{align}
\frac{1}{k} \left( 1 + \frac{1}{k - \frac{1}{2(k+1)}}\right) \leq \frac{ k^2 + 4 k + \frac{7}{2} }{k^3 + 3k^2 + \frac{1}{2} k - 2}. \label{eq:pf_lem_greedy2_ineq3}
\end{align}
Given that $k \geq 1$, inequality \eqref{eq:pf_lem_greedy2_ineq3} holds if and only if
\begin{align*}
& \left( k^2 + 2k + \frac{1}{2}\right) \left(  k^3 + 3k^2 + \frac{1}{2} k - 2 \right)  \\ 
&\qquad \leq k \left( k^2 + k - \frac{1}{2} \right) \left( k^2 + 4 k + \frac{7}{2}  \right).
\end{align*}
And the above inequality holds if and only if $(k+1)^2 \geq 0$. Thus, \eqref{eq:pf_lem_greedy2_ineq3} is satisfied if $k \geq 1$. This further implies that inequalities \eqref{eq:firm_k_plus_1}-\eqref{eq:pf_lem_greedy2_ineq2} are all satisfied. Hence, inequality \eqref{eq:cond_dec_SW} also holds for $k+1$. This completes the proof by induction.

\subsection{Proof of Lemma \ref{lem:pf_greedy3}} \label{pf:lem:pf_greedy3}
\noindent 
Let $n_k = |F_1 (\Ecal_k) |$. For the ease of exposition, we assume that the set $F_1 (\Ecal_k)$ is given by
$$F_1 (\Ecal_k) = \{1, 2, \dots, n_k-1, g_k\}.$$
However, it is straightforward to generalize the proof to the case in which $F_1 (\Ecal_k)$ is any subset of $\{1, 2, \dots, g_k\}$ satisfying $n_k = |F_1 (\Ecal_k) |$ and $\max F_1 (\Ecal_k) = g_k$.

The remainder proof proceeds in two parts. In Part 1, we provide a closed-form expression for $\SW \left( q^{\rm NE} (\Ecal_k, C^{\theta}), C^{\theta} \right)$, and show that it is piecewise quadratic in $\theta$. In Part 2, we show that given the assumption stated in this lemma, $\SW \left( q^{\rm NE} (\Ecal_k, C^{\theta}), C^{\theta} \right)$ is strictly decreasing in $\theta$ for $c_{\min} \leq \theta \leq c_{g_k}$.

\vspace{.05in}

\noindent
\emph{Part 1:} We first show that when $\theta = c_{g_k}$, we have that $q^{\rm NE}_{i1} (\Ecal, C^{\theta} ) > 0$ for all $i \in F_1 (\Ecal_k)$. First recall that 
$$\SW \left( q^{\rm NE} (\Ecal_k, C), \, C \right) > \SW  ( q^{\rm NE} (\widetilde{\Ecal}_k, C) , \, C  ).$$ 
It follows that  
$q^{\rm NE}_{g_k 1} (\Ecal_k, C) > 0.$
If this is not the case, then Nash equilibrium remains unchanged after the removal of producer $g_k$.
This implies that $\SW \left( q^{\rm NE} (\Ecal_k, C), \, C \right) = \SW  ( q^{\rm NE} (\widetilde{\Ecal}_k, C) , \, C  )$, which is a contradiction. 
Recall that $c_i \leq c_{g_k}$ for all $i \in F_1 (\Ecal_k)$. In combination with the fact that $q^{\rm NE}_{g_k 1} (\Ecal_k, C) > 0$, this implies that $q^{\rm NE}_{i1} (\Ecal, C^{\theta} ) > 0$ for all $i \in F_1 (\Ecal_k)$.


Assume that when $\theta = c_{\min}$, the vector $q^{\rm NE} (\Ecal_k, C^{\theta} )$ includes $n_{\min}$ strictly positive entries. Since $|\Ecal_k| = n_k$, we must have that $n_{\min} \leq n_k$. 
We define a collection of subsets $\Theta_{n_{\min}}, \dots, \Theta_{n_k}$ of the set $[c_{\min}, c_{g_k} ]$ according to
\begin{align*}
\Theta_{\ell} & =   \left[ \alpha_1 +  \sum_{r=1}^{\ell - 2} (\alpha_1 - c_{r} ) - \ell (\alpha_1 - c_{{\ell - 1}} ), \right. \\
&   \qquad \left. \alpha_1 +  \sum_{r=1}^{ \ell - 1} (\alpha_1 - c_{r} ) - (\ell + 1) (\alpha_1 - c_{{ \ell}} ) \right) \bigcap \left[ c_{\min}, c_{g_k} \right],
\end{align*}
for $\ell = n_{\min}, \dots, n_k - 1$, and 
\begin{align*}
\Theta_{n_k} = \left[ c_{\min}, c_{g_k} \right] \setminus \bigcup_{\ell = n_{\min}}^{n_k-1} \Theta_{\ell}.
\end{align*}
One can check that for any $\theta \in \mathrm{int}( \Theta_{\ell}) $, $\ell \in \{n_{\min}, \dots, n_k\}$, we have
$$q^{\rm NE}_{i1} (\Ecal_k, C^{\theta}) > 0, \ \text{for } i = 1, \dots, \ell - 1, \ \text{and } i = g_k.$$
That is, the vector $q^{\rm NE}  (\Ecal_k, C^{\theta}) $ contains $\ell$ strictly positive entries.
Additionally, given that the vector $q^{\rm NE} (\Ecal_k, C^{\theta} )$ includes $n_{\min}$ strictly positive entries when $\theta = c_{\min}$, we can show that
\begin{align*}
\bigcup_{\ell = n_{\min}}^{n_k} \Theta_{\ell} = \left[ c_{\min}, c_{g_k} \right], \quad  \text{and} \quad \Theta_{\ell_1} \bigcap \Theta_{\ell_2} = \emptyset
\end{align*}
for any $\ell_1, \ell_2 \in \{n_{\min}, \dots, n_k\}$ satisfying $\ell_1 \neq \ell_2$.\footnote{The collection of sets $\{\Theta_{n_{\min}}, \dots, \Theta_{n_k}\}$ might not be a partition of $[c_{\min}, c_{g_k}]$, as some of these sets can be empty.}

If follows from similar arguments as in the proof of Lemma \ref{lem:pf_greedy1} that $\SW \left( q^{\rm NE} (\Ecal_k, C^{\theta}), C^{\theta} \right)$ admits the following closed-form expression
\begin{align*}
&\SW \left( q^{\rm NE} (\Ecal_k, C^{\theta}), C^{\theta} \right) = \frac{(\alpha_1 - \theta )^2 + \sum_{r=1}^{ \ell - 1} (\alpha_1 - c_{r})^2}{\beta_1} \\
&\qquad  - \frac{2 \ell + 3 }{2} \frac{\left( \alpha_1 - \theta + \sum_{r=1}^{ \ell - 1} (\alpha_1 - c_{r}) \right)^2}{(\ell + 1)^2 \beta_1}, \quad \text{for } \ \theta \in \Theta_{\ell}.
\end{align*}
We remark that $\SW \left( q^{\rm NE} (\Ecal_k, C^{\theta}), C^{\theta} \right)$ is a piecewise quadratic function of $\theta$ that is continuous in $\theta$ for $\theta \in [c_{\min}, c_{g_k} ]$, and continuously differentiable in $\theta$ for $\theta \in {\rm int} (\Theta_{\ell} )$, $\ell \in \{ n_{\min}, \dots, n_k \}$.

\vspace{.05in}

\noindent \emph{Part 2:} We show that $\SW \left( q^{\rm NE} (\Ecal_k, C^{\theta}), C^{\theta} \right)$ is strictly monotonically decreasing in $\theta$ for $\theta \in [c_{\min}, c_{g_k} ]$. 
First recall that the union of the intervals $\Theta_{n_{\min}}, \dots, \Theta_{n_k}$ satisfies
$$\bigcup_{\ell = n_{\min}}^{n_k} \Theta_{\ell} = \left[ c_{\min}, c_{g_k} \right].$$
Additionally,
$\SW \left( q^{\rm NE} (\Ecal_k), C^{\theta} \right)$ is continuous in $\theta$ for $\theta \in [c_{\min}, c_{g_k} ]$. 
Thus, in order to show that $ \SW \left( q^{\rm NE} (\Ecal_k), C^{\theta} \right)$ is  strictly decreasing in $\theta$ on $[c_{\min}, c_{g_k} ]$, it suffices to show that 
$$\frac{\partial}{\partial \theta} \SW \left( q^{\rm NE} (\Ecal_k), C^{\theta} \right)   < 0, $$
for $ \theta \in {\rm int} (\Theta_{\ell})$, $\ell = n_{\min}, \dots, n_k$.



For $\theta \in {\rm int} (\Theta_{\ell})$, we have the following closed-form expression for $\partial \SW \left( q^{\rm NE} (\Ecal_k), C^{\theta} \right) / \partial \theta$:
\begin{align*}
&\frac{\partial}{\partial \theta} \SW \left( q^{\rm NE} (\Ecal_k), C^{\theta} \right) \\
& \qquad = \frac{ (2 \ell + 3) \sum_{r=1}^{\ell - 1} (\alpha_1 - c_{r}) - (2 \ell^2 + 2 \ell - 1) (\alpha_1 - \theta) }{(\ell  +1)^2 \beta_1}.
\end{align*}
Thus, for each $\theta \in {\rm int} (\Theta_{\ell})$, $\partial \SW \left( q^{\rm NE} (\Ecal_k), C^{\theta} \right) / \partial \theta < 0$  if the following inequality is satisfied:
\begin{align}
\alpha_1 - \theta > \frac{2 \ell + 3 }{2 \ell^2 + 2 \ell - 1} \sum_{r=1}^{\ell - 1} (\alpha_1 - c_{r}). \label{eq:cond_neg_deriv}
\end{align}

We first show that $\partial \SW \left( q^{\rm NE} (\Ecal_k), C^{\theta} \right) / \partial \theta < 0$ for $\theta \in  {\rm int} (\Theta_{n_k})$. 
Recall that $\SW \left( q^{\rm NE} (\Ecal_k, C), \, C \right) > \SW  ( q^{\rm NE} (\widetilde{\Ecal}_k, C) , \, C  )$. It follows from Lemma \ref{lem:pf_greedy1} that the following inequality is satisfied
\begin{align}
\alpha_1 - c_{g_k} > \frac{1}{n_k} \left( 1 + \frac{1}{n_k - \frac{1}{2(n_k+1)}} \right) \sum_{r=1}^{n_k - 1} (\alpha_1 - c_{r}). \label{eq:cond_dec_SW_Ek}
\end{align}
It follows from inequality \eqref{eq:cond_dec_SW_Ek} that for each $\theta  \in  {\rm int} (\Theta_{n_k})$, we have
\begin{align*}
\alpha_1 - \theta > &\alpha_1 - c_{g_k} > \frac{1}{n_k} \left( 1 + \frac{1}{ n_k - \frac{1}{2(n_k + 1)}} \right) \sum_{r=1}^{n_k - 1} (\alpha_1 - c_{r}) \\
> & \frac{2 n_k + 3 }{2 n_k^2 + 2 n_k - 1 } \sum_{r=1}^{n_k - 1} (\alpha_1 - c_{r}).
\end{align*}
Thus, $\partial \SW \left( q^{\rm NE} (\Ecal_k), C^{\theta} \right) / \partial \theta < 0$ for $\theta \in  {\rm int} (\Theta_{n_k})$. 


Next, we show that $\partial \SW \left( q^{\rm NE} (\Ecal_k), C^{\theta} \right) / \partial \theta < 0$ for $\theta \in  {\rm int} (\Theta_{\ell})$, $\ell = n_{\min}, \dots, n_k - 1$. 
Recall that $c_1 \leq \cdots \leq c_{n_k-1} \leq c_{g_k} < \alpha_1$. 
It follows from a combination of Lemma \ref{lem:pf_greedy2} and inequality \eqref{eq:cond_dec_SW_Ek} that
\begin{equation}
\begin{split}
&\alpha_1 - c_{{\ell}}  >\frac{1}{\ell} \left( 1 + \frac{1}{ \ell - \frac{1}{2(\ell + 1)}} \right) \sum_{r=1}^{\ell - 1} (\alpha_1 - c_{r})
\end{split} \label{eq:cond_dec_SW_Ek_subset}
\end{equation}
for $\ell = n_{\min}, \dots, n_k-1$.
Inequality \eqref{eq:cond_dec_SW_Ek_subset} implies that 
\begin{equation}
\begin{split}
\ell \left( \alpha_1 - c_{{\ell}} \right) 
> &\left( 1 + \frac{1}{\ell - \frac{1}{2( \ell + 1)}} \right) \sum_{r=1}^{\ell - 1} (\alpha_1 - c_{r}) \\
> &\sum_{r=1}^{\ell - 1} (\alpha_1 - c_{r})
\end{split} \label{eq:pf_greedy3_bound1}
\end{equation}
for $\ell = n_{\min}, \dots, n_k-1$.
It follows from inequality \eqref{eq:pf_greedy3_bound1} that
the following chain of inequalities are satisfied for $\theta \in {\rm int} (\Theta_{\ell})$, $\ell = n_{\min}, \dots, n_k-1$:
\begin{align*}
\theta < &\alpha_1 +  \sum_{r=1}^{ \ell - 1} (\alpha_1 - c_{r} ) - (\ell + 1) (\alpha_1 - c_{{\ell}} ) \\
= &c_{{\ell}} +  \sum_{r=1}^{\ell - 1} (\alpha_1 - c_{r} ) - \ell (\alpha_1 - c_{{\ell}} ) < c_{{ \ell}}.
\end{align*}
The above inequality, in combination with inequality \eqref{eq:cond_dec_SW_Ek_subset}, provides the following lower bound on $\alpha_1 - \theta$ for $\theta \in {\rm int} (\Theta_{\ell} )$:
\begin{align*}
\alpha_1 - \theta > &\alpha_1 - c_{{\ell}}
> \frac{1}{\ell} \left( 1 + \frac{1}{ \ell - \frac{1}{2(\ell + 1)}} \right) \sum_{r=1}^{\ell - 1} (\alpha_1 - c_{r}) \\
>&\frac{2 \ell + 3 }{2 \ell^2 + 2 \ell - 1 } \sum_{r=1}^{\ell - 1} (\alpha_1 - c_{r}).
\end{align*}
Thus, $\partial \SW \left( q^{\rm NE} (\Ecal_k), C^{\theta} \right) / \partial \theta < 0$ for $\theta \in {\rm int} (\Theta_{\ell})$, $\ell = n_{\min}, \dots, n_k-1$. This completes the proof.

\section{Search Costs}

The model and analysis presented thus far considers production costs but not other costs which can also be on the consumer side, and may be monetary, e.g., transportation and communication, or non-monetary, e.g., search costs. Platform designs center around lowering entry costs for producers, oftentimes leading to increased production and competition, and lowering search costs for consumers. In this section, we present a simple search cost model which highlights the ability of the discriminatory access design to also consider search costs, and the distinction between open access and discriminatory access designs. 

As we are working in the networked Cournot setting where demand is aggregated, we must use a simple model for search costs. We define a search cost for each consumer market $j$, $r_j$, on the consumers as a product of their consumer surplus (taking care of the simple assuring assumption that markets are rational and that search costs can never be more than their surplus) and a discount factor $f(n)$ which monotonically increases with the number of firms $n$, and lastly, a market-specific parameter $\theta$ highlighting potentially different market segments with differing search costs, defined as follows: 
$$
r_j = \theta f(n_j) CS_j, ~ 0\le \theta \le 1,
$$
where function $f$ has the following properties:
$$
f(1) = 0, \lim_{n\rightarrow \infty} f(n) = 1, f ~\text{monotonically increasing}
$$


This search cost model ensures that (i) search cost penalties can never exceed consumer surplus---an assumption that search cost can never cause the market to receive negative utility, (ii) differentiation can be made through $\theta$ for low and high search cost participants, and that (iii) search cost increases in the number of firms exposed to - the effort or cost has to increase as the number of choices increase. 

\begin{remark}
Since controlled allocation platforms essentially make all decisions on allocation, the search cost is $0$ for consumers but continue to maintain its previous negative result from Section 4, implying that search costs has no effect on controlled allocation platforms.  
\end{remark}

In light of the difficulty to analyze without a further parametric form of $f(n)$, we assume hereafter that $f(n)$ assumes the form of $f(n) = \frac{n-1}{n+1}$, fulfilling the preceding properties we listed.

Open access designs, by definition, presents all possible choices to consumers, allowing them to make production and allocation decisions, but also thereby incurring large search costs. In the following theorem, we highlight the impact of search costs on open access platforms. 

\begin{theorem}\label{thm:openaccess-search}
	Open Access Platforms with search costs defined with $f(n)=\frac{n-1}{n+1}$ have worst case efficiency loss $\Omega(n)$. 
\end{theorem}

On the other hand, referring to Corollary \ref{cor:nine}, where the price of anarchy is $\frac{4}{3}$ at $n=1$ (which coincidentally incurs zero search cost) for discriminatory access platform designs, they maintain its worst case constant bound price of anarchy under linear cost functions and search costs. 

Prior to this section, it seems that the performance guarantees attained by open access and discriminatory access platform designs are similar. This section presents a preliminary look at the impact of search cost on platform design, in particular highlighting the potentially post-search unbounded worst case efficiency loss result for open access platforms, and the flexibility of the discriminatory access platform design to consider search costs. It remains important to study this further, e.g., to find out how to optimize network design with search costs, or how computation or other costs involved may affect the performance of discriminatory access designs too.

\proof{Proof of Theorem \ref{thm:openaccess-search}}
	In the case of symmetric linear costs from the previous section, recall that consumer surplus can be written: 
	$$
	CS^{NE}_j = \frac{(\alpha_j-c_1)^2n^2}{2\beta_j(n+1)^2}
	$$
	and in this case, the overall penalty with $n$ firms can be written: 
	$$
	p_j = \theta \frac{(\alpha_j-c_1)^2n^2}{2\beta_j(n+1)^2} f(n), ~ 0\le \theta \le 1
	$$
	In this scenario, we can again make the (tight) price of anarchy explicit for the case with search costs where $\max_j \alpha_j >c$, i.e. when the solution is non-degenerate: 
	$$
	\rho(F\times M,\overline{C}) \le \frac{(n+1)^2}{n^2(1-\theta f(n))+2n}
	$$

We can already see in the above example that search costs can bring significant changes to the price of anarchy bounds we obtain. For example, when $\theta=0$, then we obtain the original bound for open access with symmetric costs firms, suggesting that the larger $n$ is, the better the bound. On the other hand, when $\theta=1$, as $n$ grows large, $\theta f(n) \rightarrow 1$, and our price of anarchy bound becomes linear in $n$, the number of firms. 

\endproof

\endproof

%% file: main.bbl
\begin{thebibliography}{68}
\providecommand{\natexlab}[1]{#1}
\providecommand{\url}[1]{\texttt{#1}}
\providecommand{\urlprefix}{URL }

\bibitem[{Abolhassani et~al.(2014)Abolhassani, Bateni, Hajiaghayi, Mahini,
  \protect\BIBand{} Sawant}]{abolhassani2014network}
Abolhassani M, Bateni MH, Hajiaghayi M, Mahini H, Sawant A (2014) Network
  cournot competition. \emph{International Conference on Web and Internet
  Economics}, 15--29 (Springer).

\bibitem[{Abreu \protect\BIBand{} Manea(2012)}]{AbreuManea2012}
Abreu D, Manea M (2012) Bargaining and efficiency in networks. \emph{Journal of
  Economic Theory} 147(1):43--70.

\bibitem[{Afeche et~al.(2018)Afeche, Liu, \protect\BIBand{}
  Maglaras}]{afeche2018ride}
Afeche P, Liu Z, Maglaras C (2018) Ride-hailing networks with strategic
  drivers: The impact of platform control capabilities on performance .

\bibitem[{Akbarpour et~al.(2017)Akbarpour, Li, \protect\BIBand{}
  Oveis~Gharan}]{akbarpour2017thickness}
Akbarpour M, Li S, Oveis~Gharan S (2017) Thickness and information in dynamic
  matching markets Available at SSRN: https://ssrn.com/abstract=2394319.

\bibitem[{Alijani et~al.(2017)Alijani, Banerjee, Gollapudi, Kollias,
  \protect\BIBand{} Munagala}]{alijani2017two}
Alijani R, Banerjee S, Gollapudi S, Kollias K, Munagala K (2017) Two-sided
  facility location. \emph{arXiv preprint arXiv:1711.11392} .

\bibitem[{Alsabah et~al.(2019)Alsabah, Bernard, Capponi, Iyengar,
  \protect\BIBand{} Sethuraman}]{alsabahcapacity}
Alsabah H, Bernard B, Capponi A, Iyengar G, Sethuraman J (2019) Multiregional
  oligopoly with capacity constraints. \emph{SSRN Electronic Journal}
  \urlprefix\url{http://dx.doi.org/10.2139/ssrn.3280688}.

\bibitem[{Anshelevich \protect\BIBand{} Sekar(2015)}]{anshelevich2015price}
Anshelevich E, Sekar S (2015) Price competition in networked markets: How do
  monopolies impact social welfare? \emph{Web and Internet Economics}, 16--30
  (Springer).

\bibitem[{Armstrong(2006)}]{armstrong2006competition}
Armstrong M (2006) Competition in two-sided markets. \emph{The RAND Journal of
  Economics} 37(3):668--691.

\bibitem[{Ashlagi et~al.(2018)Ashlagi, Burq, Jaillet, \protect\BIBand{}
  Manshadi}]{ashlagi2018matching}
Ashlagi I, Burq M, Jaillet P, Manshadi V (2018) On matching and thickness in
  heterogeneous dynamic markets. \emph{Available at SSRN 3067596} .

\bibitem[{Banerjee et~al.(2016)Banerjee, Freund, \protect\BIBand{}
  Lykouris}]{DBLP:journals/corr/Banerjee0L16}
Banerjee S, Freund D, Lykouris T (2016) Multi-objective pricing for shared
  vehicle systems. \emph{CoRR} abs/1608.06819,
  \urlprefix\url{http://arxiv.org/abs/1608.06819}.

\bibitem[{Banerjee et~al.(2017)Banerjee, Gollapudi, Kollias, \protect\BIBand{}
  Munagala}]{banerjee2017segmenting}
Banerjee S, Gollapudi S, Kollias K, Munagala K (2017) Segmenting two-sided
  markets. \emph{Proceedings of the 26th International Conference on World Wide
  Web}, 63--72.

\bibitem[{Banerjee et~al.(2015)Banerjee, Riquelme, \protect\BIBand{}
  Johari}]{banerjee2015pricing}
Banerjee S, Riquelme C, Johari R (2015) Pricing in ride-share platforms: A
  queueing-theoretic approach Available at SSRN:
  https://ssrn.com/abstract=2568258.

\bibitem[{Berry et~al.(1999)Berry, Hobbs, Meroney, O'Neill, \protect\BIBand{}
  Stewart~Jr}]{berry1999understanding}
Berry CA, Hobbs BF, Meroney WA, O'Neill RP, Stewart~Jr WR (1999) Understanding
  how market power can arise in network competition: a game theoretic approach.
  \emph{Utilities Policy} 8(3):139--158.

\bibitem[{Bimpikis et~al.(2016)Bimpikis, Candogan, \protect\BIBand{}
  Saban}]{bimpikis2016spatial}
Bimpikis K, Candogan O, Saban D (2016) Spatial pricing in ride-sharing networks
  Available at SSRN: https://ssrn.com/abstract=2868080.

\bibitem[{Bimpikis et~al.(2014)Bimpikis, Ehsani, \protect\BIBand{}
  Ilkilic}]{bimpikis2014cournot}
Bimpikis K, Ehsani S, Ilkilic R (2014) Cournot competition in networked
  markets. \emph{EC}, 733.

\bibitem[{Bose et~al.(2014)Bose, Cai, Low, \protect\BIBand{}
  Wierman}]{bose2014role}
Bose S, Cai DW, Low S, Wierman A (2014) The role of a market maker in networked
  cournot competition. \emph{53rd IEEE Conference on Decision and Control},
  4479--4484 (IEEE).

\bibitem[{Boudreau(2010)}]{boudreau2010open}
Boudreau K (2010) Open platform strategies and innovation: Granting access vs.
  devolving control. \emph{Management Science} 56(10):1849--1872.

\bibitem[{Cai et~al.(2017)Cai, Bose, \protect\BIBand{} Wierman}]{cai2017role}
Cai D, Bose S, Wierman A (2017) On the role of a market maker in networked
  cournot competition. \emph{arXiv preprint arXiv:1701.08896} .

\bibitem[{Chawla et~al.(2007)Chawla, Hartline, \protect\BIBand{}
  Kleinberg}]{chawla2007algorithmic}
Chawla S, Hartline JD, Kleinberg R (2007) Algorithmic pricing via virtual
  valuations. \emph{Proceedings of the 8th ACM conference on Electronic
  commerce}, 243--251 (ACM).

\bibitem[{Chawla et~al.(2010)Chawla, Hartline, Malec, \protect\BIBand{}
  Sivan}]{chawla2010multi}
Chawla S, Hartline JD, Malec DL, Sivan B (2010) Multi-parameter mechanism
  design and sequential posted pricing. \emph{Proceedings of the forty-second
  ACM symposium on Theory of computing}, 311--320 (ACM).

\bibitem[{Chawla \protect\BIBand{} Roughgarden(2008)}]{chawla2008bertrand}
Chawla S, Roughgarden T (2008) Bertrand competition in networks.
  \emph{Algorithmic Game Theory}, 70--82 (Springer).

\bibitem[{Chen(2017)}]{chen2017thrown}
Chen JY (2017) Thrown under the bus and outrunning it! the logic of didi and
  taxi drivers’ labour and activism in the on-demand economy. \emph{New Media
  \& Society} 1461444817729149.

\bibitem[{Chen et~al.(2016)Chen, Mislove, \protect\BIBand{}
  Wilson}]{chen2016empirical}
Chen L, Mislove A, Wilson C (2016) An empirical analysis of algorithmic pricing
  on amazon marketplace. \emph{Proceedings of the 25th International Conference
  on World Wide Web}, 1339--1349 (International World Wide Web Conferences
  Steering Committee).

\bibitem[{Chen \protect\BIBand{} Wilson(2017)}]{chen2017observing}
Chen L, Wilson C (2017) Observing algorithmic marketplaces in-the-wild.
  \emph{ACM SIGecom Exchanges} 15(2):34--39.

\bibitem[{Dinerstein et~al.(2018)Dinerstein, Einav, Levin, \protect\BIBand{}
  Sundaresan}]{dinerstein2018consumer}
Dinerstein M, Einav L, Levin J, Sundaresan N (2018) Consumer price search and
  platform design in internet commerce. \emph{American Economic Review}
  108(7):1820--59.

\bibitem[{Einav et~al.(2015)Einav, Kuchler, Levin, \protect\BIBand{}
  Sundaresan}]{einav2015assessing}
Einav L, Kuchler T, Levin J, Sundaresan N (2015) Assessing sale strategies in
  online markets using matched listings. \emph{American Economic Journal:
  Microeconomics} 7(2):215--247.

\bibitem[{Elliott(2015)}]{Elliott2015}
Elliott M (2015) Inefficiencies in networked markets. \emph{American Economic
  Journal: Microeconomics} 7(4):43--82,
  \urlprefix\url{http://dx.doi.org/10.1257/mic.20130098}.

\bibitem[{Evans \protect\BIBand{} Schmalensee(2005)}]{evans2005industrial}
Evans DS, Schmalensee R (2005) The industrial organization of markets with
  two-sided platforms. Technical report, National Bureau of Economic Research.

\bibitem[{Evans \protect\BIBand{} Gawer(2016)}]{evans2016rise}
Evans P, Gawer A (2016) The rise of the platform enterprise: A global survey.
  \urlprefix\url{http://epubs.surrey.ac.uk/811201/}.

\bibitem[{Fang et~al.(2018)Fang, Huang, \protect\BIBand{}
  Wierman}]{fang2018loyalty}
Fang Z, Huang L, Wierman A (2018) Loyalty programs in the sharing economy:
  Optimality and competition. \emph{arXiv preprint arXiv:1805.03581} .

\bibitem[{Forbes(2018)}]{forbesreport}
Forbes (2018) Global 2000: The world’s largest public companies.
  \urlprefix\url{https://www.forbes.com/global2000/#50eb23d5335d}.

\bibitem[{Gross \protect\BIBand{} Acquisti(2003)}]{gross2003balances}
Gross B, Acquisti A (2003) Balances of power on ebay: Peers or unequals.
  \emph{Workshop on Economics of Peer-to-peer systems} (Citeseer).

\bibitem[{Guzm{\'a}n(2011)}]{guzman2011price}
Guzm{\'a}n CL (2011) Price competition on network. Technical report.

\bibitem[{Hagiu \protect\BIBand{} Wright(2015)}]{hagiu2015multi}
Hagiu A, Wright J (2015) Multi-sided platforms. \emph{International Journal of
  Industrial Organization} 43:162--174.

\bibitem[{Hobbs et~al.(2000)Hobbs, Metzler, \protect\BIBand{}
  Pang}]{hobbs2000strategic}
Hobbs BF, Metzler CB, Pang JS (2000) Strategic gaming analysis for electric
  power systems: An mpec approach. \emph{IEEE transactions on power systems}
  15(2):638--645.

\bibitem[{Holmberg \protect\BIBand{} Philpott(2014)}]{holmberg2014supply}
Holmberg P, Philpott A (2014) Supply function equilibria in transportation
  networks .

\bibitem[{Hui et~al.(2016)Hui, Saeedi, Shen, \protect\BIBand{}
  Sundaresan}]{hui2016reputation}
Hui X, Saeedi M, Shen Z, Sundaresan N (2016) Reputation and regulations:
  evidence from ebay. \emph{Management Science} 62(12):3604--3616.

\bibitem[{Ilkilic(2009)}]{ilkilic2009cournot}
Ilkilic R (2009) Cournot competition on a network of markets and firms.
  \emph{FEEM Working Paper} Available at SSRN:
  https://ssrn.com/abstract=1443216.

\bibitem[{Johari \protect\BIBand{} Tsitsiklis(2005)}]{johari2005efficiency}
Johari R, Tsitsiklis JN (2005) Efficiency loss in cournot games.
  \emph{Technical Report}
  \urlprefix\url{http://web.mit.edu/jnt/www/Papers/R-05-cournot-tr.pdf}.

\bibitem[{Kanoria \protect\BIBand{} Saban(2017)}]{kanoria2017facilitating}
Kanoria Y, Saban D (2017) Facilitating the search for partners on matching
  platforms: Restricting agents' actions. \emph{Columbia Business School
  Research Paper No. 17-74} Available at SSRN:
  https://ssrn.com/abstract=3004814.

\bibitem[{Koutsoupias \protect\BIBand{} Papadimitriou(1999)}]{KP99}
Koutsoupias E, Papadimitriou CH (1999) Worst-case equilibria. \emph{Symposium
  on Theoretical Aspects of Computer Science}, 404--413.

\bibitem[{Kreps \protect\BIBand{} Scheinkman(1983)}]{kreps1983quantity}
Kreps DM, Scheinkman JA (1983) Quantity precommitment and bertrand competition
  yield cournot outcomes. \emph{The Bell Journal of Economics} 326--337.

\bibitem[{Lahiri \protect\BIBand{} Ono(1988)}]{lahiri1988helping}
Lahiri S, Ono Y (1988) Helping minor firms reduces welfare. \emph{The economic
  journal} 98(393):1199--1202.

\bibitem[{Le~Cadre(2018)}]{le2018efficiency}
Le~Cadre H (2018) On the efficiency of local electricity markets under
  decentralized and centralized designs: a multi-leader stackelberg game
  analysis. \emph{Central European Journal of Operations Research} 1--32.

\bibitem[{Li et~al.(2012)Li, Ng, Trayer, \protect\BIBand{}
  Liu}]{li2012automated}
Li Y, Ng BL, Trayer M, Liu L (2012) Automated residential demand response:
  Algorithmic implications of pricing models. \emph{IEEE Transactions on Smart
  Grid} 3(4):1712--1721.

\bibitem[{Liang et~al.(2017)Liang, Schuckert, Law, \protect\BIBand{}
  Chen}]{liang2017superhost}
Liang S, Schuckert M, Law R, Chen CC (2017) Be a “superhost”: The
  importance of badge systems for peer-to-peer rental accommodations.
  \emph{Tourism management} 60:454--465.

\bibitem[{Lin \protect\BIBand{} Bitar(2019)}]{lin2019structural}
Lin W, Bitar E (2019) A structural characterization of market power in electric
  power networks. \emph{IEEE Transactions on Network Science and Engineering} .

\bibitem[{Liu \protect\BIBand{} Wu(2007)}]{liu2007impacts}
Liu Y, Wu FF (2007) Impacts of network constraints on electricity market
  equilibrium. \emph{IEEE Transactions on Power Systems} 22(1):126--135.

\bibitem[{Lu et~al.(2018)Lu, Frazier, \protect\BIBand{} Kislev}]{lu2018surge}
Lu A, Frazier PI, Kislev O (2018) Surge pricing moves uber's driver-partners.
  \emph{Proceedings of the 2018 ACM Conference on Economics and Computation},
  3--3 (ACM).

\bibitem[{Luca(2017)}]{luca2017designing}
Luca M (2017) Designing online marketplaces: Trust and reputation mechanisms.
  \emph{Innovation Policy and the Economy} 17(1):77--93.

\bibitem[{Luo et~al.(1996)Luo, Pang, \protect\BIBand{}
  Ralph}]{luo1996mathematical}
Luo ZQ, Pang JS, Ralph D (1996) \emph{Mathematical programs with equilibrium
  constraints} (Cambridge University Press).

\bibitem[{Ma et~al.(2018)Ma, Fang, \protect\BIBand{} Parkes}]{ma2018spatio}
Ma H, Fang F, Parkes DC (2018) Spatio-temporal pricing for ridesharing
  platforms. \emph{arXiv preprint arXiv:1801.04015} .

\bibitem[{Mankiw \protect\BIBand{} Whinston(1986)}]{mankiw1986free}
Mankiw NG, Whinston MD (1986) Free entry and social inefficiency. \emph{The
  RAND Journal of Economics} 48--58.

\bibitem[{Mas-Colell et~al.(1995)Mas-Colell, Whinston, Green
  et~al.}]{mas1995microeconomic}
Mas-Colell A, Whinston MD, Green JR, et~al. (1995) \emph{Microeconomic theory},
  volume~1 (Oxford university press New York).

\bibitem[{Motalleb et~al.(2017)Motalleb, Eshraghi, Reihani, Sangrody,
  \protect\BIBand{} Ghorbani}]{motalleb2017game}
Motalleb M, Eshraghi A, Reihani E, Sangrody H, Ghorbani R (2017) A
  game-theoretic demand response market with networked competition model.
  \emph{Power Symposium (NAPS), 2017 North American}, 1--6 (IEEE).

\bibitem[{Nava(2015)}]{Nava2015}
Nava F (2015) Efficiency in decentralized oligopolistic markets. \emph{Journal
  of Economic Theory} 157:315--348.

\bibitem[{Nguyen(2015)}]{nguyen2015coalitional}
Nguyen T (2015) Coalitional bargaining in networks. \emph{Operations Research}
  63(3):501--511.

\bibitem[{Nosko \protect\BIBand{} Tadelis(2015)}]{nosko2015limits}
Nosko C, Tadelis S (2015) The limits of reputation in platform markets: An
  empirical analysis and field experiment. Technical report, National Bureau of
  Economic Research.

\bibitem[{Parker \protect\BIBand{} Van~Alstyne(2017)}]{parker2017innovation}
Parker G, Van~Alstyne M (2017) Innovation, openness, and platform control.
  \emph{Management Science} .

\bibitem[{Rochet \protect\BIBand{} Tirole(2003)}]{rochet2003platform}
Rochet JC, Tirole J (2003) Platform competition in two-sided markets.
  \emph{Journal of the european economic association} 1(4):990--1029.

\bibitem[{Rysman(2009)}]{rysman2009economics}
Rysman M (2009) The economics of two-sided markets. \emph{Journal of Economic
  Perspectives} 23(3):125--43.

\bibitem[{Scheiber(2017)}]{scheiber2017uber}
Scheiber N (2017) How uber uses psychological tricks to push its drivers'
  buttons. \emph{The New York Times} 2.

\bibitem[{Suzumura \protect\BIBand{} Kiyono(1987)}]{suzumura1987entry}
Suzumura K, Kiyono K (1987) Entry barriers and economic welfare. \emph{The
  Review of Economic Studies} 54(1):157--167.

\bibitem[{Tadelis(2016)}]{tadelis2016reputation}
Tadelis S (2016) Reputation and feedback systems in online platform markets.
  \emph{Annual Review of Economics} 8:321--340.

\bibitem[{Weyl(2010)}]{weyl2010price}
Weyl EG (2010) A price theory of multi-sided platforms. \emph{The American
  Economic Review} 100(4):1642--1672.

\bibitem[{Wilson(2008)}]{wilson2008supply}
Wilson R (2008) Supply function equilibrium in a constrained transmission
  system. \emph{Operations research} 56(2):369--382.

\bibitem[{Xu et~al.(2017)Xu, Cai, Bose, \protect\BIBand{}
  Wierman}]{xu2017efficiency}
Xu Y, Cai D, Bose S, Wierman A (2017) On the efficiency of networked
  stackelberg competition. \emph{Information Sciences and Systems (CISS), 2017
  51st Annual Conference on}, 1--6 (IEEE).

\bibitem[{Yao et~al.(2008)Yao, Adler, \protect\BIBand{} Oren}]{yao2008modeling}
Yao J, Adler I, Oren S (2008) Modeling and computing two-settlement
  oligopolistic equilibrium in a congested electricity network.
  \emph{Operations Research} 56(1):34--47.

\end{thebibliography}
